\documentclass[11pt,a4paper]{article}

\usepackage{amsmath, amssymb,amsfonts}  
\usepackage{physics} 
\usepackage{layout}
\usepackage{graphicx}      
\usepackage{amssymb}
\usepackage{caption}
\usepackage{subcaption}
\usepackage{float}
\usepackage{import}
\usepackage{xcolor}
\usepackage{transparent}
\usepackage{cite}

\usepackage[utf8]{inputenc}       
\usepackage[T1]{fontenc}
\usepackage[hidelinks]{hyperref}

\newlength{\textlength}
\newlength{\margin}
\newcommand*\Diff[1]{\mathop{}\!\mathrm{d^#1}}

\newcommand\blfootnote[1]{%
  \begingroup
  \renewcommand\thefootnote{}\footnote{#1}%
  \addtocounter{footnote}{-1}%
  \endgroup
}

\begin{document}

\thispagestyle{empty}

\begin{flushright}
{\small
MITP-25-072}
\end{flushright}

\vspace{-0.5cm}

\begin{center}
\Large\bf\boldmath
Bubble velocities in local equilibrium from a pseudopotential
\unboldmath
\end{center}

\vspace{-0.2cm}

\begin{center}
Martin M\"unzenberg$^{*}$\blfootnote{$^*$ mmuenzen@students.uni-mainz.de} and Carlos Tamarit$^\dagger$\blfootnote{$^\dagger$ ctamarit@uni-mainz.de}\\
\vskip0.4cm

{\it PRISMA+ Cluster of Excellence \& Mainz Institute for Theoretical Physics,\\
Johannes Gutenberg-Universität Mainz, 55099 Mainz, Germany}
\end{center}

\begin{abstract}
We present a new method to estimate terminal bubble velocities during first-order phase transitions in a plasma in local equilibrium. The method relies on calculating the extrema of a modified potential function for the scalar field undergoing the transition. The shape of this function, which we refer to as the ``pseudopotential'', changes with the wall velocity, and if the dependence of the fluid temperature on scalar gradients is weak --- which is confirmed to hold with high accuracy in concrete examples --- the difference in pseudopotential between two appropriate extrema gives the net outward pressure acting on the bubble wall. It then follows that the correct terminal bubble velocities are those that lead to degenerate minima in the pseudopotential. This allows to compute bubble velocities without having to solve the equation of motion of the scalar field, and in contrast to other methods this can be done without relying on simplified equations of state for the plasma or without choosing a specific ansatz for the scalar field profile. We illustrate the method in a singlet extension of the Standard Model, computing the net outward pressure as a function of the wall velocity. We confirm the dip in outward pressure found in the literature for hybrid bubbles, which implies that stationary deflagrations are stable, while their detonation counterparts are unstable.
\end{abstract}

\section{Introduction}
A current scenario beyond the standard model (SM) of certain interest is the occurrence of cosmological first-order phase transitions in the early universe. During its thermal evolution, the early universe is believed to have experienced multiple cosmological phase transitions, with the two most significant examples being the QCD phase transition and the electroweak phase transition. However, both phase transitions are a crossover in the SM \cite{Aoki:2006we,Kajantie:1996mn,Kajantie:1996qd}. In a crossover transition there is no discontinuity in the thermodynamic quantities, and no true phase boundary forms. Consequently, no propagating phase front develops and the crossover transitions lack in out-of-equilibrium dynamics. A first order phase transition, on the other hand, proceeds through the formation of bubbles of true vacuum, which expand and convert the surrounding space into the favoured phase. This motivates the study of first order phase transitions: Firstly, they could be a candidate to solve the baryon asymmetry problem, creating the Sakharov conditions \cite{Sakharov:1967dj} necessary for electroweak baryogenesis \cite{Kuzmin:1985mm,Rubakov:1996vz,Morrissey:2012db,Garbrecht:2018mrp}. Secondly, there are scenarios in which first order phase transitions could lead to a stochastic background of gravitational waves generated by colliding vacuum bubbles \cite{Witten:1984rs,Kosowsky:1991ua,Kamionkowski:1993fg}. These gravitational waves will be potentially detectable by next-generation interferometers like LISA (Laser Interferometer Space Antenna) \cite{LISA:2017pwj,Caprini:2019egz,Caprini:2015zlo,Athron:2023xlk}.

Both phenomena mentioned before are strongly influenced by the velocity of the expanding bubbles. For example, baryogenesis requires a relatively slow wall velocity to allow the CP asymmetry generated at the bubble wall to diffuse into the symmetric phase ahead of the wall, where sphaleron interactions can convert this asymmetry into a baryon number asymmetry \cite{Heckler:1994uu}. In contrast, the production of gravitational wave signals relies on high energies and therefore on faster bubble walls, particularly on bubble walls becoming nearly luminal \cite{Hindmarsh:2017gnf}. Therefore, predicting the bubble wall velocity is essential for making theoretical models to observable phenomena.

The expansion velocity of the bubble wall is determined by the total vacuum pressure acting on the wall, which consists of a driving force accelerating the bubble and a friction force opposing its motion. While it was widely accepted that  the friction force arose solely from out-of-equilibrium effects \cite{Moore:1995si}, recent studies have established that the friction force also exists in local thermal equilibrium (LTE), where it is governed by hydrodynamic backreaction effects that give rise to temperature changes across the bubble wall \cite{Konstandin:2010dm,BarrosoMancha:2020fay,Balaji:2020yrx,Ai:2021kak,Krajewski:2024gma}. This has led to an intense study of the role of hydrodynamic effects and the validity of the LTE approximation. It is generally expected that estimates of the bubble wall velocities using the LTE approximation should provide an upper bound on the wall velocity, as non-equilibrium effects will lead to additional friction \cite{Eriksson:2025owh}. While the presence of a backreaction force even in LTE suggests that bubbles reaching a constant, subluminal velocity should be generally expected, runaway behaviour is still possible if the maximum backreaction force is below the driving force. Such considerations have led to an LTE-based criterion for runaway behaviour \cite{Ai:2024shx}, which has general validity in regions of parameter space in which the maximum backreaction force can be larger than the dominant non-equilibrium effects. Most of the studies are based on looking for static solutions to the equations governing the dynamics of the scalar fields and the plasma in a planar approximation. This captures the regions near the bubble wall of bubbles moving with constant velocity. Far in front of the bubble wall, in which the scalar field takes constant values and its dynamics are irrelevant, one should recover a fluid at rest at the nucleation temperature, which requires additional fluid dynamics. It is well known that there are three types of hydrodynamic solutions (see figure \ref{fig:three_profiles}), which in order of increasing bubble wall speed are: Deflagrations, where the fluid is perturbed in front of the bubble wall; hybrids, with the fluid perturbed in front and behind; and detonations, in which the fluid is perturbed behind the wall. Going beyond constant velocity bubbles, recent simulations of time-dependent bubble expansions have shown that the static configurations predicted in the LTE approximation are not always reached, and in particular static detonations were never obtained \cite{Krajewski:2024gma}.

With this caveat in mind, in this article we focus on static solutions in LTE. To find these, in principle one has to solve the hydrodynamic equations together with the scalar equation of motion \cite{Ignatius:1993qn,Espinosa:2010hh}. As this method requires nontrivial computations, many studies simplify the analysis by modelling the thermal plasma with a parametrisation that allows eschewing the solution of the scalar equation of motion. With the scalar going to a constant away from the bubble wall, one can focus on these regions and use stress-energy conservation to impose matching constraints across the wall \cite{Espinosa:2010hh}. In the case of LTE, one has an additional constraint coming from entropy conservation \cite{Ai:2021kak}. If the stress-energy tensor of the plasma is known in both phases, the constraints are enough to calculate the fluid temperatures and velocities on both sides of the wall, together with the bubble wall velocity. Initially, the method relying on the entropy constraint was applied using simplifying assumptions on the equation of state. In Ref.~\cite{Ai:2021kak} the plasma was modelled with the so-called bag equation of state \cite{Chodos:1974je}, while Ref.~\cite{Ai:2023see}  used a more general parametrisation introduced in Refs.~\cite{Giese:2020rtr,Giese:2020znk}. More recently, the {\tt WallGo code} \cite{Ekstedt:2024fyq} implemented the LTE calculation with entropy matching for arbitrary models in the case of deflagration and hybrids. The popular bag model treats the plasma in each phase as an ideal gas with constant effective degrees of freedom and a fixed vacuum energy difference, allowing for an analytic treatment of the hydrodynamic matching conditions \cite{Ignatius:1993qn,Espinosa:2010hh}. While it captures the key thermodynamic features of the phase transition, such as the latent heat and the pressure jump across the wall, it neglects effects from mass scales other than the temperature, which can lead to quantitative differences in the predicted wall velocity. Even when using more general parametrisations as in Refs.~\cite{Ai:2023see,Giese:2020rtr,Giese:2020znk}, an issue remains: once the parametrisation is matched to the underlying microscopical physical model --- which can be done in terms of the effective potential at finite temperature --- the parameters are not constant but background- and temperature-dependent. Hence, a consistent description for a particular particle physics model requires making sure that the chosen parameters are consistent with the solutions of the matching equations.
The previous caveats do not apply to calculations formulated directly in terms of thermodynamic quantities obtained from the effective potential, allowing for a description based on a general equation of state, as realised for instance in {\tt WallGo}.

An alternative route to determine the wall velocity is to directly solve the scalar equation of motion. An intermediate approach relies on approximating the wall profile with a hyperbolic tangent ansatz depending on a few parameters. The latter can be determined by considering moments of the equation of motion evaluated on the ansatz and minimising the result  \cite{Laurent:2022jrs,Ekstedt:2024fyq}. This procedure does not yield exact solutions and introduces systematic uncertainties, as the ansatz does not reflect the true shape of the field profile determined by the underlying potential. Among the results obtained with this approach, a notable find is that the backreaction pressure opposing the driving force, when considered as a function of the wall velocity, has a peak at the minimum velocity allowed for detonations, known as the Jouguet velocity \cite{Laurent:2022jrs}. This suggests that if bubbles nucleate at low velocities and expand, they will encounter maximum backreaction while expanding as deflagrations or hybrids. Hence, it is expected that only static deflagration or hybrid solutions will be reached in practice \cite{Ai:2023see}, in accordance with the aforementioned results of the time-dependent simulations in Ref.~\cite{Krajewski:2024gma}.

In this paper, we develop an alternative method for finding bubble wall velocities in LTE, which avoids the need to solve the scalar equation of motion explicitly and does not rely on assumptions about the field profile or the underlying potential. While such features are shared with the LTE calculations in {\tt WallGo}, which makes use of entropy matching, the approach introduced in this paper relies on defining a quantity with a direct physical interpretation that remains useful beyond the static regime. This quantity is a generalised potential or ``pseudopotential'', and it contains information such as the net outward pressure acting on bubble configurations interpolating between minima of the scalar potential. While static configurations can be obtained by demanding a null outward pressure --- which offers an alternative to the entropy matching constraint --- the study of configurations with nonzero outward pressure can give additional information about issues of stability, which cannot be directly inferred with the entropy matching method. The method is built up starting with the balance of forces in LTE as discussed in \cite{Ai:2021kak}. We note that, whenever one can neglect the dependence of the temperature profile on gradients of the scalar field --- which in the examples considered in this paper works at the per mille level --- one can define a generalised potential or ``pseudopotential'', which takes the wall velocity $v_w$ as an input parameter and has the following properties. First, the field values that minimise the standard potential also extremise the pseudopotential, though the corresponding extremum need not be a minimum. Second, the difference between the values of the pseudopotential evaluated at the extrema can be directly interpreted as the net pressure acting on the bubble wall. The latter is zero for bubbles expanding with constant velocity, so that $v_w$ can be estimated by requiring degenerate minima in the pseudopotential. In this sense, the method is complementary to entropy-conservation approaches such as used in {\tt WallGo}: besides providing an efficient route to the static solutions, the pseudopotential itself has a direct physical interpretation and can also be used to analyse configurations away from the stationary points. Furthermore, as we will show, the method also allows us to treat detonation configurations within LTE, including cases where no stationary solutions are realised.

In this article we introduce the method and demonstrate its application and accuracy in a singlet extension of the SM. We calculate the net pressure as a function of the wall velocity and show that it can be zero for both deflagrations and detonations, reflecting the existence of static solutions of both types. The predictions for $v_w$ using the pseudopotential will be shown to accurately match those obtained by solving the scalar equation of motion. By plotting the net pressure as a function of the wall velocity, we confirm the existence of a peak in the backreaction force, as found in Ref.~\cite{Laurent:2022jrs} with the $\tanh$ method. However, in our case the peak is found for hybrid solutions with $v_w$ near the speed of sound, as opposed to hybrid solutions with $v_w$ nearing the Jouguet velocity. The slope of the total outward pressure as a function of velocity is negative for deflagrations and positive for detonations, suggesting that while the static deflagrations are stable, static detonations are unstable \cite{Ai:2023see}, which again connects with the findings in Ref.~\cite{Krajewski:2024gma}.

The structure of the paper is outlined as follows: In section~\ref{sec:hydrodynamic_equations} we review the equations governing the dynamics of the scalar field and the plasma, as well as the deflagration, detonation and hybrid solutions. Section~\ref{sec:pseudopotential} introduces the pseudopotential that captures the net pressure on the bubble wall as long as one can ignore the impact of field gradients on the temperature. In section~\ref{sec:example_model} we present the model used in the remainder of this work, namely an extension of the SM with $N$ complex singlets. In section~\ref{sec:validity} the validity of the pseudopotential method is tested by explicitly comparing the resulting bubble wall velocities with those obtained by solving the scalar equation of motion. In section~\ref{sec:results} we analyse the net pressure as a function of the wall velocity and show how it provides insight into the stability of static solutions. Finally, our findings are summarised in section~\ref{sec:conclusion}.

\section{Hydrodynamic Equations}
\label{sec:hydrodynamic_equations}

In this section we review the standard equations governing bubble propagation,  following from stress-energy conservation \cite{Ignatius:1993qn,Espinosa:2010hh}. The scalar equation of motion can also be derived from functional methods in non-equilibrium quantum field theory, which predict additional quantum corrections \cite{Ai:2025bjw}. These novel terms are formally of higher order in a loop expansion and will not be considered here. With this in mind, we consider a scalar field $\phi$ whose stress-energy tensor is defined by
\begin{align}
T^{\mu \nu}_{\phi} &= \partial^{\mu} \phi \partial^{\nu} \phi - g ^{\mu \nu} \left( \frac{1}{2} \partial_{\rho} \phi \partial^{\rho} \phi - V(\phi) \right) ,
\label{eq:stress_energy_scalar}
\end{align}
where $V(\phi)$ is the vacuum potential of the scalar field including loop corrections.
We treat $\phi$ as a background field, interacting with all the other particle content in the theory. At high temperatures the particles form a plasma described by a stress-energy tensor
\begin{align}
T^{\mu \nu}_{p} &= \sum_i \int \frac{\Diff3 p}{(2\pi)^3 E_i} p^\mu p^\nu f_i(p,x),
\label{eq:stress_energy_plasma_1}
\end{align}
with $f_i(p,x)$ being the distribution functions summed up over all the plasma's degrees of freedom. In the early universe one has high density and frequent particle interactions, such that the velocity and temperature gradients are usually kept small and one can model the plasma by a perfect fluid as a leading-order approximation. The stress-energy tensor in eq.~\eqref{eq:stress_energy_plasma_1} then becomes 
\begin{align}
T^{\mu \nu}_{p} &= (\rho + p) u^{\mu} u^{\nu} - \eta^{\mu \nu} p = \omega u^{\mu} u^{\nu} - \eta^{\mu \nu} p ,
\label{eq:stress_energy_plasma_2}
\end{align}
with $u^\mu$ being the fluid four velocity and $\rho$, $p$ and  $\omega = p + \rho$ being the energy density, pressure and enthalpy of the fluid respectively. The four-velocity vector $u^\mu = \gamma(1,v^1,v^2,v^3)$ of the plasma can be expressed in terms of the Lorentz factor $\gamma=1/\sqrt{1-v^2}$ and the three-velocity $v^i$ with $i=1,2,3$ and magnitude $v = \sqrt{\sum_i(v^i)^2}$. In LTE, all the information about $\rho$ and $p$ is beautifully encoded by the effective potential including thermal corrections, $V(\phi,T)$, which can be computed in finite-temperature quantum field theory. The thermal corrections to the potential correspond to the free-energy density, which is related to the pressure by
\begin{align}\label{eq:pressure_V}
p=-V(\phi,T)+V(\phi)\equiv -V^T(\phi,T).                                                                                                                                                                                                              \end{align}
Additionally, standard thermodynamic identities yield
\begin{align}\label{eq:rho_V}
 \rho =  T \frac{\partial p}{\partial T} -p= - T \frac{\partial V^T(\phi,T)}{\partial T}+V^T(\phi,T).
\end{align}
The total energy momentum tensor of the 'wall-plasma' system is then defined as a sum of eqs.~\eqref{eq:stress_energy_scalar} and \eqref{eq:stress_energy_plasma_2}:
\begin{align}
T^{\mu \nu} = T^{\mu \nu}_{\phi} + T^{\mu \nu}_{p}.
\end{align}
Covariant conservation of the energy-momentum tensor yields 
\begin{align}
 \nabla_\mu T ^{\mu \nu}= 0 \ .
 \label{eq:cov_conservation}
\end{align}

From now on we neglect the expansion of the universe by assuming that the Hubble rate is negligible compared to the relevant physical scales, allowing us to perform our analysis in flat Minkowski spacetime. Then the covariant derivative is replaced by an ordinary one and eq.~\eqref{eq:cov_conservation} together with eqs.~\eqref{eq:pressure_V} and \eqref{eq:rho_V} lead to
\begin{align}
\label{eq:scalar_eom}
\Box \phi + \frac{\partial}{\partial \phi} ( V(\phi,T)) &= 0,\\
\label{eq:pressure_equations}
\partial_{\mu} (\omega u^{\mu} u^{\nu} - \eta^{\mu \nu} p) + \frac{\partial p}{\partial \phi} \partial^{\nu} \phi &= 0.
\end{align}
Notice that the first equation is simply the scalar equation of motion at finite temperature, with the potential $ V(\phi,T)$ including thermal corrections.
In the following, we assume that for high temperatures the potential $ V(\phi,T)$ has a single minimum at a vacuum expectation value (VEV) $\langle\phi\rangle=\phi_+=0$. We will refer to the phase characterised by this VEV as the ``symmetric phase''. At smaller temperatures, an extra minimum appears at $\langle\phi\rangle=\phi_-\neq0$, corresponding to a new phase which will be denoted as ``broken phase''. The two minima are assumed to be separated by an energy barrier, leading to a first-order phase transition proceeding through bubble nucleation. The transition starts when the universe reaches a nucleation temperature $T_{\textrm{nuc}}$, for which the rate of bubble formation exceeds the rate of expansion of the universe. This temperature is typically lower than the critical temperature $T_c$, at which the symmetric and broken phases are degenerate in free energy.

For bubbles that are sufficiently large and propagate with constant speed, the bubble wall structure near the interface is effectively static in the wall's frame, and any time evolution can be neglected. In addition, if the bubble radius is much larger than the wall thickness, the curvature effects become irrelevant and the wall can be locally approximated as planar. Strictly speaking, the planar/static approximation fails for large $|z|$; the underlying assumption is that the thickness of the bubble wall --- i.e. the extent of the spatial range with significant field gradients --- is much smaller than the length scales associated with variations of the fluid temperature and velocity very far away from the wall. Under these approximations, in a region in which the wall advances in the positive $z$-direction and the fluid  velocity is $v^x = v^y = 0, v^z \equiv v$, one finds that eqs.~\eqref{eq:scalar_eom} and \eqref{eq:pressure_equations} imply \cite{Ignatius:1993qn}
\begin{align}
& - \phi^{\prime \prime} (z) + \frac{\partial}{\partial \phi} V(\phi, T) = 0 \ , \label{eq:scalar_eom_planar} \\
& \omega \gamma^2 v^2 + \frac{1}{2} (\phi ' (z))^2 - V(\phi,T) = c_1  ,\quad \omega \gamma^2 v = c_2 \ , \label{eq:wall_profile_equation}
\end{align}
with $c_1,c_2$ being constants. The relations \eqref{eq:pressure_V}, \eqref{eq:rho_V} imply that eqs.~\eqref{eq:scalar_eom_planar}, \eqref{eq:wall_profile_equation} depend only on the field $\phi$ and its derivatives, the velocity $v$ and temperature $T$.  The field ahead of the wall is expected to relax to the VEV of the high-temperature phase, which yields the boundary conditions $\phi(z)=0, \ z \to \infty$ and $\phi^\prime (z) =0, \ z \to \infty$.
Behind the wall, we expect the field to relax to a nonzero constant,
$\phi^\prime (z) =0, \ z \to - \infty$. These boundary conditions together with eqs.~\eqref{eq:scalar_eom_planar} and \eqref{eq:wall_profile_equation} form in practice a system that can be solved via e.g. a shooting method. First, by considering eqs.~\eqref{eq:wall_profile_equation} far in front of the wall, with $\phi=\phi'=0$, the constants $c_1,c_2$ can be traded for the temperature $T_+$ and the velocity $v_+$ in front of the wall. Given then a choice of $v_+,T_+$ one can start at $z\rightarrow-\infty$,  setting $\phi'=0$ and scanning the initial values of $\phi(z=-\infty)\equiv \phi_-$ until a unique solution to eqs.~\eqref{eq:scalar_eom_planar} and \eqref{eq:wall_profile_equation} is found with $\phi,\phi'$ going to zero at $z \to \infty$. Having thus started with some values of $T_+,v_+$, one ends up predicting the values of all relevant quantities across the wall, including in particular the field $\phi_-(v_+,T_+)$,  the temperature $T_-(v_+,T_+)$ and velocity $v_-(v_+,T_+)$  far behind the wall.
From now on we identify quantities in front/behind the wall, as measured in the rest frame of the bubble wall, with the subscripts $+/-$. An additional constraint comes from the expectation that far behind the bubble the physical field configurations should asymptotically settle into the VEV $\phi_-$ of the broken phase. The latter is a minimum of the potential, satisfying $\frac{\partial}{\partial \phi}\left. V(\phi, T)\right|_{\phi_-,T_-} = 0$, which together with eq.~\eqref{eq:scalar_eom_planar} yields an additional condition: $\phi^{\prime \prime}(z)=0, \ z \to -\infty$. As a consequence of the additional constraint, $T_+$ and $v_+$ cannot be chosen independently, leaving one free parameter. 

Far from the bubble wall, the planar approximation eventually stops being valid. While the scalar field can still be taken as constant in these regions, the profiles for the temperature and velocity can no longer be static. A way to understand this is that, in the limit in which the plasma particles are massless and the scalar background is constant, the plasma has no intrinsic dimensionful scale. Therefore the solutions of the hydrodynamic equations can only depend on the scale-invariant ratio $\xi=r/t$, where $r$ is the distance from the centre of the bubble and $t$ the time since nucleation \cite{Gyulassy:1983rq,Espinosa:2010hh}. Hence the solutions are necessarily time-dependent, yet with a particular functional form that is referred to as ``self-similar''. We can identify $\xi$ as the velocity of a specific point in the wave profile and $v(\xi)$ as the velocity of the fluid at the point $\xi$ in the rest frame of the bubble centre. The latter matches what we refer to as the universe frame, since far behind and in front of the bubble wall the fluid is at rest. We will use both conventions interchangeably. In summary, the static solution near the wall has to be matched with self-similar solutions of the hydrodynamic equations \eqref{eq:pressure_equations} in a constant scalar background. Here we have additional boundary conditions, namely that the temperature for large $\xi$ should recover the nucleation temperature $T_{\textrm{nuc}}$, and that the fluid in the universe/plasma frame should be at rest. As the static solutions had a free parameter, the two additional conditions would seem to overconstrain the system, but there is an extra unknown in the problem, which is the boost needed to go from the universe frame to the bubble wall frame. The velocity of this boost is nothing but the bubble wall velocity, which ends up being calculable, as the number of unknowns matches the number of constraints.

Under the hypothesis of self-similar solutions, eq.~\eqref{eq:pressure_equations} yields two differential equations
\begin{align}
\label{eq:hydro_diff_equation1}
\frac{\xi-v}{\omega} \partial_\xi \rho - 2 \frac{v}{\xi} - (1-\gamma^2 v (\xi - v)) \partial_\xi v &= 0 \ , \\
\label{eq:hydro_diff_equation2}
\frac{1-v \xi}{\omega} \partial_\xi p - \gamma^2 (\xi-v) \partial_\xi v &= 0 \ .
\end{align}

The pressure and density are now understood to be evaluated at the constant value of the field appropriate for each phase. The solutions to these differential equations have been thoroughly studied, particularly in the simplifying case in which the pressure and temperature on each phase mimic those of a relativistic ideal gas \cite{Chodos:1974je,Espinosa:2010hh}. This is the so-called ``bag model'' with the pressure and energy densities in the symmetric and broken phases having the form\footnote{In some references such as~\cite{Espinosa:2010hh}, a vacuum energy contribution $\epsilon$ is added to $\rho_+$, and subtracted to $p_+$. In our conventions, the $\epsilon$-dependent factors correspond to the stress-energy tensor of the scalar field, eq.~\eqref{eq:stress_energy_scalar}, evaluated at $\phi=\phi_+$ with $\epsilon=V(\phi_+,T_+)$. As in this paper we have separated the fluid and scalar contributions to $T^{\mu\nu}$, we do not include the $\epsilon$ dependence in the definition of the pressure and density of the fluid.}
\begin{align}\label{eq:bag}
 p_\pm = \frac{1}{3} a_{\pm } T^4, \quad \rho_\pm = a_\pm T^4.
\end{align}

\begin{figure}[!t]
     \centering
     \captionsetup[subfigure]{labelformat=empty,font=normalsize,skip=-2pt}
     \begin{subfigure}[b]{0.31\textwidth}
         \centering
         \includegraphics[width=\textwidth]{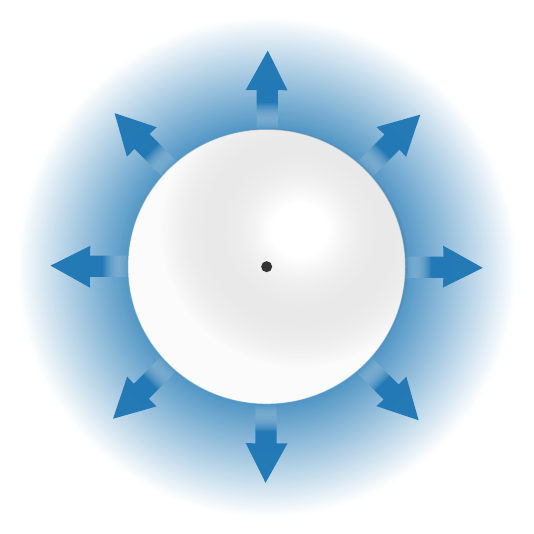}
         \caption{Deflagration}
         \label{fig:deflag_scetch}
     \end{subfigure}
     \hfill
     \begin{subfigure}[b]{0.31\textwidth}
         \centering
         \includegraphics[width=\textwidth]{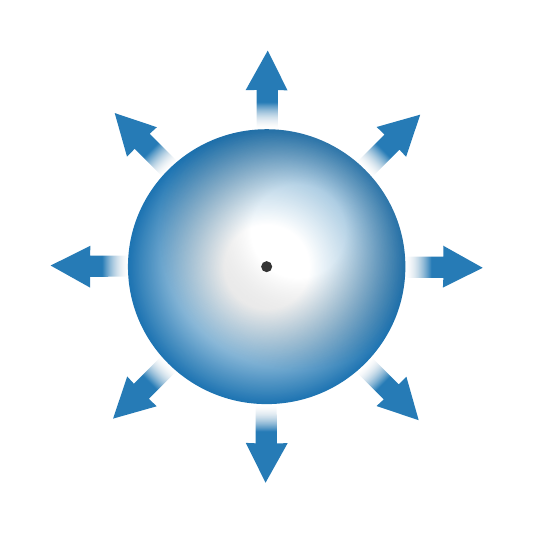}
         \caption{Detonation}
         \label{fig:deton_scetch}
     \end{subfigure}
     \hfill
     \begin{subfigure}[b]{0.31\textwidth}
         \centering
         \includegraphics[width=\textwidth]{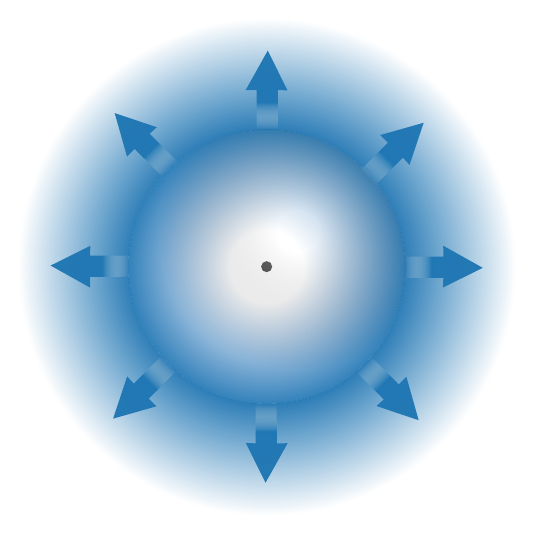}
         \caption{Hybrid}
         \label{fig:hybrid_scetch}
     \end{subfigure}
    \caption{Different types of expanding bubbles with corresponding qualitative fluid velocity profile. The sphere represents the bubble wall moving in the direction of the arrows, and the blue contour depicts regions with nonzero fluid velocity. For deflagrations one has $v_w < c_s$, while hybrids and detonations satisfy $v_w > c_s$.}
    \label{fig:three_profiles}
\end{figure}

A schematic visualisation of the solutions allowed for the bag model can be found in figure \ref{fig:three_profiles}. For small wall velocities, specifically $v_w<c_s$ with $c_s^2 = \partial_T p / \partial_T \rho $ being the speed of sound in the plasma, the fluid heats up in front of the bubble wall and moves with a velocity $v_\text{fluid}(\xi)$, while behind the wall it remains at rest. These solutions are referred to as deflagrations. When boosting to the wall frame, the fluid behind the wall --- which was at rest in the universe frame --- will be seen receding with a velocity $v_-=-v_w$. The second type of solutions, detonations, involve a rarefaction wave behind the bubble front, while the fluid in front of the wall remains unperturbed. Consequently, boosting to the wall frame gives $v_+=-v_w$. Detonation profiles occur when $v_w>c_s$. In addition to deflagrations and detonations there exists a third type of solution, called hybrids. As the name suggests, they represent a combination of detonations and deflagrations and are characterised by both a shock front ahead of the wall and a rarefaction wave behind the wall.

In detonations, the fluid gradually comes to rest behind the wall without further discontinuities. In contrast, deflagrations and hybrid solutions can feature an additional discontinuity at a shock front preceding the wall. To ensure consistency of the hydrodynamic profile, one must relate the fluid variables across this shock by imposing the conservation of the energy-momentum tensor, as expressed in eqs.~\eqref{eq:scalar_eom_planar} and \eqref{eq:wall_profile_equation}. A detailed discussion of the matching procedure for detonation, deflagration and hybrid profiles can be found in the appendix~\ref{sec:hydrodynamic_matching}.
 
Ref.~\cite{Ai:2021kak} pointed out that within LTE it is not strictly necessary to solve the scalar equation of motion. The alternative relies in matching the values of the total stress-energy tensor and the entropy current across the bubble wall. This is equivalent to solving eqs.~\eqref{eq:scalar_eom} and \eqref{eq:pressure_equations}, because the latter encode  not only stress energy conservation $\partial_\mu T^{\mu \nu}$, but in LTE they also imply conservation of the entropy current \cite{Balaji:2020yrx}, $\partial_\mu j_s^\mu=0,$ for $j_s^\mu = s u^\mu$, with $s= \mathrm{d}p/\mathrm{d}T$ being the entropy density. In the wall frame one can thus start with 7 unknowns, given by $\phi_\pm,T_\pm,v_\pm$ and the wall velocity $v_w$ parametrising the boost from the universe frame. There are three matching equations for $T^{0z}, T^{zz}$ and $s^z$ across the wall. In front of the latter the plasma should be in the symmetry phase with $\phi_+=0$, while behind the wall one should require $\phi_-$ to  minimise the potential, $\frac{\partial}{\partial \phi}V(\phi,T)|_{\phi_-,T_-}=0$. This gives 5 constraints that leave 2 free parameters out of the initial 7. It remains, however, to match the quantities $T_\pm,v_\pm$ in front and behind the bubble wall with hydrodynamic profiles that satisfy two additional constraints, namely that the temperature relaxes to $T_{\textrm{nuc}}$ far in front of the bubble, and that the fluid velocity goes to zero. This fixes all unknowns including the bubble wall velocity $v_w$. A simplified version of this method was applied for the bag equation of state in Ref.~\cite{Ai:2021kak}. With the bag parametrisation of eq.~\eqref{eq:bag}, when treating $a_\pm$ as constants, the background field dependence of $\rho$ and $p$ is hidden. As a result, the unknowns and constraints related to $\phi_\pm$ can be dropped,  leaving a system with five unknowns and five constraints, which is considerably simpler to manage. However, when matching the bag parametrisation to a given microscopic model, the coefficients $a_\pm$, which follow from the effective potential $V(\phi,T)$, are in reality temperature and field-dependent. Fixing their correct values makes it unavoidable to introduce $\phi_-$ and its associated minimisation constraint. The method of matching stress-energy and entropy in LTE has been applied beyond the bag parametrisation in Ref.~\cite{Ai:2023see}. Here, when matching to a microscopic model, one will face a similar issue as before, with the parameters characterising the system being field-dependent. Again, one has to consider $\phi_-$ and its associated minimisation condition.

In the next section, we introduce an alternative procedure to determine the wall velocity without solving the scalar equation of motion, and without making use of a simplified parametrisation of the stress-energy momentum of the plasma. The method is based on directly estimating the total pressure on the bubble wall as a function of the wall velocity, thereby providing additional physical insights.

\section{Pseudopotential as a  balance of forces}
\label{sec:pseudopotential}
The  equation of motion for the scalar field, without assuming LTE of the plasma and neglecting the quantum corrections found in Ref.~\cite{Ai:2025bjw}, is given by \cite{Moore:1995ua,Moore:1995si}
\begin{align}
\Box \phi + \frac{\partial  V(\phi,T)}{\partial \phi} + \sum_i \dv{m_i^2(\phi)}{\phi} \int \frac{\Diff3 \mathbf{p}}{(2 \pi)^3 2 E_i} \delta f_i (p,x) = 0\ .
\label{eq:Full_Scalar_EoM}
\end{align}
Here $\delta f (p,x)$ is the out-of-equilibrium contribution from the particle distribution function and the sum includes all the particle content of the theory. For a long time it was widely accepted that a friction force decelerating the bubble wall can only originate from the non-equilibrium term in \eqref{eq:Full_Scalar_EoM}.  However, hints or counterexamples to the contrary could already be found in Refs.~\cite{Moore:1995si,Ignatius:1993qn,Kurki-Suonio:1996gkq,Espinosa:2010hh}, while the first calculations specifically on LTE were carried out in Ref.~\cite{Konstandin:2010dm}, showing that LTE can  already yield subluminal deflagrations. The importance of LTE in general settings was highlighted more recently in Ref.~\cite{BarrosoMancha:2020fay}. Subsequently, Refs.~\cite{Balaji:2020yrx,Ai:2021kak} clarified that the same effect was underlying the results in previous works, with the friction-like behaviour in LTE arising from temperature gradients across the wall enforced by the hydrodynamic equations. These temperature gradients naturally generate a non-dissipative friction force which counters the acceleration of the bubble wall originated from the difference in vacuum pressure. In this way, the runaway behaviour may be averted. This balance of forces in LTE can be seen by following the derivation in \cite{Ai:2021kak}. The starting point is eq.~\eqref{eq:Full_Scalar_EoM} with $\delta f = 0$ in the static limit, corresponding to LTE and a bubble wall at rest. Assuming the wall to be planar and perpendicular to the $z$ direction,  multiplying the equation by $\mathrm{d}\phi / \mathrm{d}z$ and integrating over $z$ gives
\begin{align}
\int \mathrm{d}z \dv{\phi}{z}\bigg( \Box \phi + \frac{\partial  V(\phi,T)}{\partial \phi} \bigg) = \int \mathrm{d}z \bigg[ \dv{}{z} \bigg( \frac{1}{2} \phi^{\prime 2} \bigg) +  \dv{\phi}{z} \frac{\partial  V(\phi,T)}{\partial \phi} \bigg] = 0 \ .
\label{eq:Full_Scalar_EoM2}
\end{align}
As the scalar field is constant at large $|z|$, the first term vanishes. Applying a simple chain rule pulls apart the second term and allows us to rewrite the integral as
\begin{align}
\int \mathrm{d}z \dv{\phi}{z} \frac{\partial V (\phi, T)}{\partial \phi} = \int \mathrm{d}z \left( \dv{V(\phi, T)}{z} - \frac{\partial V(\phi, T)}{\partial T} \dv{T}{z} \right) = 0 \ .
\label{eq:balance of forces}
\end{align}
The first term on the right-hand side corresponds to the difference in the effective potential $\Delta V(\phi,T)$ between the symmetric and broken phases. It represents the total pressure difference across the bubble wall, which drives its propagation. The second term can be seen as a hydrodynamic backreaction force arising from the temperature gradient $\mathrm{d}T / \mathrm{d}z$ that slows down the bubble and prevents it from reaching luminal velocities. In a steady-state regime, where the bubble wall propagates at a constant velocity, we expect these two forces to balance each other in magnitude: The driving force is exactly counteracted by the hydrodynamic backreaction force. This concept motivates recasting the balance of forces in terms of an effective function, offering an easy and clear criterion for identifying stable bubble wall solutions.

To do so, let us first return to eqs.~\eqref{eq:scalar_eom_planar} and \eqref{eq:wall_profile_equation}. As described in section~\ref{sec:hydrodynamic_equations}, for a specific configuration of $v_+$ and $T_+$ one can determine both constants $c_1$ and $c_2$. With the latter fixed, considering now general values of $z$, eqs.~\eqref{eq:wall_profile_equation} will define $v(\phi,\phi^\prime)$ and $T(\phi,\phi^\prime)$ in terms of the field and its derivative. Plugging back the field-dependent temperature into the thermal potential leaves a function $V(\phi,\phi^\prime) = V(\phi,T(\phi,\phi^\prime))$ that depends purely on the field $\phi(z)$ and its derivative $\phi^\prime (z)$. The only remaining equation is the scalar equation of motion, but with a modified “potential” --- a term conventionally used here but somewhat misleading, as this function depends on $\phi^\prime(z)$ and thus does not align with the physical definition of a true potential. To simplify further, an approximation is made by neglecting the $\phi^\prime$-dependence in the temperature, reducing $T(\phi,\phi^\prime)$ to $T(\phi)$. This allows the construction of a purely field-dependent potential by expressing eq.~\eqref{eq:balance of forces} in terms of the newly defined potential $V(\phi,T(\phi,\phi^\prime=0))$. Starting from the left-hand side of eq.~\eqref{eq:balance of forces},  we choose to relabel the field as $\phi \rightarrow \varphi$ and change the integration variable $z$ to $\varphi$. This leads to the following function:
\begin{align}
\hat{V}(\phi) \equiv \int_0^\phi \mathrm{d} \varphi ~ \frac{\partial V}{\partial \varphi}  (\varphi, T (\varphi)) \ .
\label{eq:Pseudopotential}
\end{align}
The change of variables from $z$ to $\varphi$ is well defined, as for a bubble configuration with a field profile that is monotonic in $z$ we have an invertible relation between the scalar field $\varphi$ and $z$. We will refer to the field-dependent function in eq.~\eqref{eq:Pseudopotential} as the ``pseudopotential''. It provides an approximate framework to describe the balance of forces across the bubble wall for a specific configuration of the velocity and temperature in front of the bubble. As argued before, eq.~\eqref{eq:balance of forces} represents the balance of forces across the bubble wall. In terms of the pseudopotential, the left-hand-side of eq.~\eqref{eq:balance of forces} becomes $\hat V(\phi_+)-\hat V(\phi_-)$. Hence we conclude that
\begin{align}\label{eq:Vhat_forces}
 \Delta \hat V(\phi)\equiv\hat V(\phi_+)-\hat V(\phi_-)=\text{Driving pressure}-\text{Backreaction pressure}=\text{Net outward pressure}.
\end{align}
That is, we managed to express the balance of forces in terms of a field-dependent function. This construction works provided that the dependence of $T(\phi,\phi')$ on $\phi'$ can be neglected.

Taking things further, eq.~\eqref{eq:Pseudopotential} implies
\begin{align}
\frac{\mathrm{d}\hat V(\phi)}{\mathrm{d}\phi}=\frac{\partial V(\phi,T(\phi))}{\partial\phi},
\label{eq:Vminima}
\end{align}
so that the equation of motion, eq.~\eqref{eq:scalar_eom_planar}, can be written in terms of the total derivative of the pseudopotential:
\begin{align}
 - \phi^{\prime \prime} (z) + \frac{\mathrm{d}}{\mathrm{d} \phi} \hat{V}(\phi) = 0
 \label{eq:scalar_eom_Pseudopotential}.
\end{align}
Defining the energy function
\begin{align}
E \equiv \frac{1}{2} \left( \dv{\phi(z)}{z} \right)^2 - \hat{V}(\phi)
\label{eq:energy_function}
\end{align}
and taking the derivative with respect to $z$ gives
\begin{align}
\frac{\mathrm{d}}{\mathrm{d} z} E = \dv{\phi(z)}{z} \left( \phi^{\prime \prime} (z) - \frac{\mathrm{d}}{\mathrm{d} \phi} \hat{V} (\phi) \right) \ .
\label{eq:conserved_energy_function}
\end{align}
The term in the brackets can be identified with the scalar equation of motion in eq.~\eqref{eq:scalar_eom_Pseudopotential}. Thus the energy function is conserved if the field configuration satisfies the equation of motion. Instead of solving the second-order differential equation for $\phi(z)$ directly, the conservation of $E$ reduces the analysis significantly. If the energy function is conserved, we deduce from eq.~\eqref{eq:energy_function} and the fact, that the field derivatives must go to zero far way from the wall, that $\hat{V}(\phi)$ must have the same values in the symmetric phase $\phi_+ \equiv \phi(z \to \infty)$ and the broken phase $\phi_- \equiv \phi(z \to -\infty)$:
\begin{align}
\Delta \hat{V} = \hat{V}(\phi_+) - \hat{V} (\phi_-) = 0 \ .
\label{eq:effpressure}
\end{align}
Hence, we arrive again at the conclusion that, for a static bubble, the net pressure acting on the bubble, captured by $\Delta \hat{V}$ as in eq.~\eqref{eq:Vhat_forces}, is zero.

A further property of $\hat V(\phi)$ is that the field values of its extrema match those of the usual potential $V(\phi,T)$, a consequence of eq.~\eqref{eq:Vminima}. Then the boundary conditions for physical bubbles, that were discussed before, imply that bubble profiles interpolate between extrema of $\hat V(\phi)$. An interesting possibility is that minima of $V(\phi,T)$ do not necessarily have to  correspond to minima of $\hat V(\phi)$, but may instead appear as maxima or saddle points; we illustrate examples of this behaviour in section~\ref{sec:results}. In this case, however, if one has a maximum and a minimum of $\hat V(\phi)$, one can directly rule out a physical static configuration with $\Delta\hat V=0$ between the extrema.

\begin{figure}[!t]
\centering
\def\svgwidth{0.7\textwidth}
\hskip1cm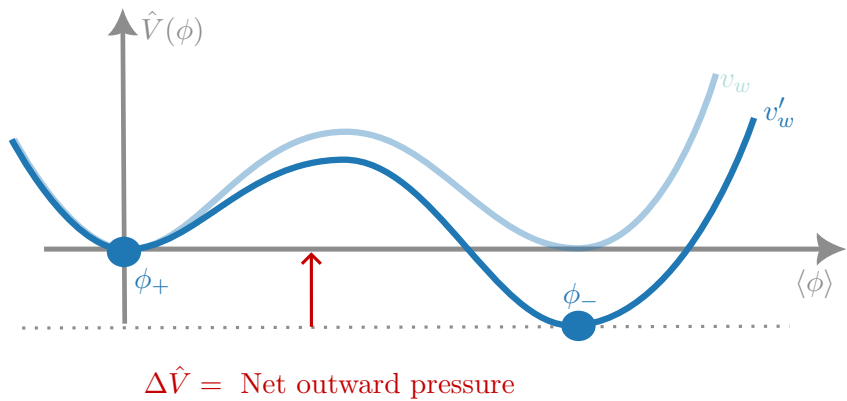
\caption{Illustration of the physical meaning of the pseudopotential. Its extrema coincide with the minima $\phi_{\pm}$ of the ordinary potential on which the field is expected to settle ahead of/behind the bubble wall. The difference of the values of $\hat V(\phi)$ at the extrema gives the net outward pressure on the bubble wall. Changing the wall velocity $v_w$ changes the pseudopotential $\hat V(\phi)$ and with it the outward pressure,  which is zero for stationary bubbles.}
\label{fig:Vhat_qualitative}
\end{figure}

The former results offer a new avenue to determine the wall velocity $v_w$ without solving the static equations of motion. The correct value of $v_w$ will be the one ensuring that eq.~\eqref{eq:effpressure} is satisfied, with $\phi_+=0$ the extremum in the symmetric phase and $\phi_-$ given by a second, nontrivial extremum of $\hat{V}(\phi)$. The pseudopotential is sensitive to the wall velocity, because the function $T(\varphi)$ used to construct $\hat{V}(\phi)$, as in eq.~\eqref{eq:Pseudopotential}, depends both on the velocity $v_+$ and the temperature $T_+$ in front of the bubble wall, which are related to $v_w$. For example, for detonations one directly has $v_w=-v_+$ (see figure \ref{fig:detonation_qualitative}). For deflagrations, one has $v_w=-v_-$ (see figure \ref{fig:deflagration_qualitative}), and $v_-$ itself can be computed given $v_+,T_+$ by evaluating the solution $v(\phi,\phi')$ of the hydrodynamic equations \eqref{eq:wall_profile_equation} for $\phi=\phi_-$ (computed by extremising $\hat V(\phi)$), and $\phi'=0$. That is, for a specific configuration of the velocity $v_+$ and temperature $T_+$, one can check whether the pseudopotential difference between the two relevant extrema satisfies $\Delta \hat{V} = 0$. If this condition holds, the configuration is associated with a physically consistent static bubble wall where the driving and friction forces are in balance, and which solves the static equations of motion. A nonzero result for $\Delta \hat{V}$  would indicate an imbalance between the driving force and the friction force, implying an unstable field configuration that does not solve the static equations of motion. Moreover, as $\Delta \hat V$ corresponds to the net outward pressure, $\Delta \hat V>0$ would correspond to accelerating bubbles, while $\Delta \hat V<0$ would indicate deceleration. The terminal velocity for bubbles that reach constant speed can then be found by scanning $v_+,T_+$ until $\Delta \hat{V} = 0$ is satisfied and the correct nucleation temperature is recovered far in front of the wall (including the effects of a shock front for deflagrations and hybrids). The two unknowns $v_+,T_+$ can thus be determined with the two constraints, and with $v_+,T_+$ fixed, the bubble wall velocity can be computed as explained before. We apply this method to an example model in the following sections. The physical meaning of the pseudopotential is illustrated in figure \ref{fig:Vhat_qualitative}.

We emphasise that the whole analysis presented above relies on the assumption that one can neglect the derivative of the field in the solution $T(\phi,\phi')$ to the hydrodynamic equations \eqref{eq:wall_profile_equation}. For a non-negligible dependence of $T$ on $\phi^\prime(z)$, $\hat V(\phi)$ would depend on both $\phi$ and $\phi'$. In that case, the energy function is no longer conserved and $\Delta\hat V = 0$ no longer provides a criterion for physical solutions; the balance of forces described earlier would therefore no longer hold. Hence, before proceeding to calculate $\Delta \hat{V}$ in our example model for different configurations, we demonstrate the validity of the approximation $T(\phi,\phi')\approx T(\phi,0)$. This will be achieved by explicitly solving the scalar equations of motion and comparing the results to those obtained by demanding $\Delta \hat{V} = 0$.

Before considering a concrete application of the proposed method to compute bubble velocities in LTE, let us  summarise its advantages with respect to other approaches in the literature. First, the method avoids the need to solve the differential equation for the scalar profile, while retaining the need to solve the algebraic equations \eqref{eq:wall_profile_equation} for $T$ and $v$. Eschewing the scalar equation of motion is done without introducing simplifying assumptions on the shape of the bubble profile  or the stress-energy momentum of the plasma. This is in contrast to commonly used approaches which either impose a $\tanh$-profile for the bubble wall or model the plasma with the bag equation of state or a generalisation thereof. With our method, there is no need to assume a shape for the bubble profile, and the pressure and density of the plasma are derived from the thermal effective potential in LTE, computed from first principles.

\section{Example model}
\label{sec:example_model}
In order to illustrate our method we shall consider a simple extension of the SM in which the electroweak phase transition is rendered first-order, thus proceeding by bubble nucleation. For this we extend the SM by $N$ complex scalar singlets $s$ in a multiplet $S = \{s_1,s_2,\dots,s_n\}$ with couplings invariant under U($N$)-transformations. The multiplet couples to the SU(2) Higgs doublet
\begin{align}
H = \frac{1}{\sqrt{2}}\begin{pmatrix}
		\chi_1+ i \chi_2 \\
		h + i \chi_4
	\end{pmatrix} \
	\label{eq:higgs_doublet_raw}
\end{align}
and the phase transition is assumed to only change the background of the neutral Higgs component $h$. The Lagrangian of the model involves the following tree-level potential for the scalar fields:
\begin{equation}
\mathcal{L} \supset -V_{\textrm{tree}}=- \frac{\left(m_H^2\right)^2}{2\lambda} - m_H^2H^\dagger H - \frac{\lambda}{2} (H^\dagger H)^2 - m_S^2 S^\dagger S - \frac{\lambda_S}{2} (S^\dagger S)^2 - \lambda_{H S} S^\dagger S H^\dagger H,
\label{eq:potential_SM_plus_scalar}
\end{equation}
where the first term  is simply there to ensure that the potential is  zero at the classical zero temperature minimum
\begin{align}
\langle h \rangle = v = \sqrt{-\frac{2 m_H^2}{\lambda}} \ .
\end{align}

To formulate the hydrodynamic equations we use the finite temperature effective potential up to one loop order, evaluated on a nonzero background for $h$:
\begin{align}
V(h,T) = V_\text{tree}(h) + V_1(h) + V_1^T(h,T).
\end{align}
The zero temperature one-loop correction $V_1(h)$ in Landau gauge is explicitly given by
\begin{align}
V_1(h) = \sum_i \frac{(-1)^{2s_i} n_i m_i^4(h)}{64\pi^2} \left[ \log \left( \frac{m_i^2(h)}{\mu^2}\right) - c_i \right].
\end{align}
In the previous equation, $i$ labels the different particle species, while $s_i$ denotes the corresponding spin and $n_i$ is the associated number of degrees of freedom, where each real scalar accounts for one d.o.f., each Weyl fermion 2 d.o.f., and gauge bosons 3 d.o.f.. The mass of each particle species in the Higgs background is denoted as $m_i(h)$, and additionally one has $c_i=3/2$ for scalars and fermions and $c_i=5/6$ for gauge bosons. In the subsequent calculations the renormalisation scale $\mu$ will be fixed at $\mu^2 = m_W^2$. The finite temperature contribution $ V_1^T(h,T) $ up to one-loop order is obtained as (see \cite{Quiros:1999jp} for a review)
\begin{align}
V_1^ T (h,T) = \frac{T^4}{2 \pi^2} \sum_i n_i J_{B/F} \left(\frac{m_i^2(h)}{ T^2}\right),
\label{eq:thermal_potential_sum}
\end{align}
with $J_B$/$J_F$ representing the thermal bosonic/fermionic loop functions, given below together with their corresponding asymptotic expansions for small arguments:
\begin{align}\label{eq:JFBs}\begin{aligned}
J_{B}(x^2)=&\,\int_0^\infty\,\mathrm{d}y\;y^2\log[1- e^{-\sqrt{x^2+y^2}}]=-\frac{\pi^4}{45}+\frac{\pi^2}{12}\,x^2-\frac{\pi}{6}x^{3}+{\cal O}\left(x^4\right),\\
J_{F}(x^2)=&\,\int_0^\infty\,\mathrm{d}y\; y^2\log[1+e^{-\sqrt{x^2+y^2}}]=\frac{7\pi^4}{360}-\frac{\pi^2}{24}\,x^2+{\cal O}\left(x^4\right).
\end{aligned}\end{align}
We account for all the SM particle species plus the $N$ complex scalar singlets. For the fermions, we neglect Yukawa couplings other than those of the top and bottom quarks, with all lighter fermions treated as massless. Notice that massless particles, although not contributing to the background-field-dependent part of the effective potential, will still contribute to the temperature dependent part. This is important to obtain the correct pressure (see eq.~\eqref{eq:pressure_V}) which is sensitive to all the relativistic degrees of freedom. Using the small argument expansions in eq.~\eqref{eq:JFBs} (which apply at high temperature), eq.~\eqref{eq:thermal_potential_sum} can be determined up to terms of order T as follows:
\begin{equation}
\begin{split}
V_1^T(h,T) = & - \frac{1}{90} \pi^2 T^4 \left(g_{\ast,SM} + 2N\right)\\
& + T^2 \bigg[\bigg( \frac{y_t^2 }{8} + \frac{y_b^2 }{8} +\frac{3 g_1^2}{160} + \frac{3 g_2^2}{32} + \frac{\lambda}{8} + \frac{N \lambda_{HS}}{24} \bigg) h^2 \ + \frac{m_H^2}{6} + \frac{N m_S^2}{12} \bigg]\\
& +  \frac{T}{12 \pi} \bigg[ - \frac{3}{4} \Big( g_2^2~ h^2 \Big)^{3/2}  - \frac{3}{8} \bigg( \bigg(\frac{3 }{5} g_1^{2}+ g_2^2 \bigg) h^2 \bigg) ^{3/2} - 3 \bigg( \frac{\lambda}{2}  h^2+ m_H^2 \bigg) ^{3/2} \\ 
& \qquad \qquad \qquad \qquad \qquad \qquad - \bigg( \frac{3}{2} \lambda h^2+ m_H^2 \bigg)^{3/2} - 2N  \bigg( \frac{\lambda_{HS}}{2} h^2 + m_S^2 \bigg)^{3/2}  \; \bigg]  \ .
\end{split}
\label{eq:thermal_potential_high_temp}
\end{equation}
In this equation, $g_{\ast,SM} = 106.75$ is the number of effective relativistic degrees of freedom in the SM plasma, $y_t$ and $y_b$ denote the top and bottom-quark Yukawa couplings, $g_2$ is the usual SU(2) gauge coupling, and $g_1$ is the hypercharge gauge coupling defined in the GUT normalisation convention, related to the canonically normalised U(1) coupling $g^{\prime}$ via $g^{\prime} = {\textstyle \sqrt{3/5}}\,g_1$~\cite{Georgi:1974yf}. We fix $m_S^2/m_W^2 = 0.0625$ and vary the number of complex singlets $N$ and the quartic coupling $\lambda_{HS}$ to investigate their impact on the bubble wall velocity. The term proportional to $T$ in eq.~\eqref{eq:thermal_potential_high_temp} gives rise to a potential barrier between the two phases, which triggers a first-order phase transition. The strength of the transition can be controlled by changing $N$ and $\lambda_{HS}$.

\section{Exact treatment vs pseudopotential method}
\label{sec:validity}
In this section, we analyse the accuracy of the estimates of wall velocities in LTE obtained from the pseudopotential method --- i.e. imposing eq.~\eqref{eq:effpressure} --- by comparing them with the velocities extracted from solving the static scalar equation of motion
\begin{align}
-h^{\prime \prime}(z) + \dv{}{h} V(h,T(h,h^\prime)) = 0,
\label{eq:differential_equation}
\end{align}
with boundary conditions $h(z) \rightarrow 0, z \to \infty $,  $h^{\prime}(z)\rightarrow 0, z \to \pm \infty$ as well as $h^{\prime \prime}(z) \rightarrow 0, ~ z \to -\infty$. The potential $V(h,T(h,h^\prime))$ incorporates the full $h^\prime$ dependence of the temperature function, as derived from eqs.~\eqref{eq:wall_profile_equation}. Throughout this work, we vary the number of scalar degrees of freedom and the portal coupling in order to probe different phase transition strengths, considering $N \in [4,10]$ and $\lambda_{HS} \in [0.70, 0.95]$.

Among the inputs  required for finding consistent bubble profiles, the only quantity that must be determined is the nucleation temperature $T_\text{nuc}$. This temperature should be recovered far in front of the wall. $T_\text{nuc}$ is determined by solving for static critical bubbles, computing their associated three dimensional action $S_3[h,T]$, and demanding $S_3[h,T_{\textrm{nuc}}]/T_{\textrm{nuc}} \approx 140$. This ensures that the probability of nucleating a bubble inside a Hubble patch (with the nucleation rate per unit volume going as $\Gamma \sim e^{-S_3/T}$) becomes of order one  \cite{Quiros:1999jp}.

Once $T_\text{nuc}$ is determined, we can start matching the static solution to eqs.~\eqref{eq:wall_profile_equation} and \eqref{eq:differential_equation}  (or the corresponding values of $h_+,h_-$ inferred with the pseudopotential method) with hydrodynamic profiles extending  away from the wall and recovering the nucleation temperature far in front of the bubble. We will refer to this as ``hydrodynamic matching''; a full description is provided in  appendix~\ref{sec:hydrodynamic_matching}. For both deflagrations or detonations, the matching starts by choosing $v_+,T_+$. One may then either solve the static scalar equation \eqref{eq:differential_equation} with boundary conditions  $h(z) \rightarrow 0, z \to \infty $,  $h^{\prime}(z)\rightarrow 0, z \to \pm \infty$, which allows to extract $h_-$; or, within the pseudopotential method, identifying $h_-$ with the location of the minimum away from $h=0$ by imposing
\begin{align}
\left. \dv{}{h} \hat{V}(h)\right|_{h=h_-} = 0.
\end{align}
The former admits a solution only if the broken phase of $V(h,T(h,h^\prime))$ has lower free energy than the symmetric one; if this condition is not met, eq.~\eqref{eq:differential_equation} and the boundary conditions cannot be simultaneously fulfilled. Once $h_-$ is known, eqs.~\eqref{eq:wall_profile_equation} fix the hydrodynamic variables throughout the wall, and evaluating them far behind the wall in the broken phase $(h=h_-,\,h'=0)$ directly gives $v_-$ and $T_-$. For either detonations (with $v_+ = -v_w$) or deflagrations (with $v_- = -v_w$) one ends up with a fixed wall velocity, which allows all quantities to be expressed in the fluid frame. These can then be matched to the corresponding hydrodynamic profiles in front of the wall (for deflagrations) and behind the wall (for detonations) by solving eqs.~\eqref{eq:hydro_diff_equation1} and \eqref{eq:hydro_diff_equation2}. Possible discontinuity fronts, such as a shock front in a deflagration, are handled by relating the fields across the front using identities that follow from eqs.~\eqref{eq:wall_profile_equation}. For the initially chosen $v_+,T_+$, this procedure fixes all the relevant values of the field,  temperature and velocity, including in particular $h_-$ and the temperature far ahead of the bubble and possible deflagration front. It then remains to scan $v_+,T_+$ until the two remaining constraints are satisfied, namely that the temperature for $z\rightarrow\infty$ matches the nucleation temperature, and an additional condition that depends on whether one solves the scalar equation of motion or uses the pseudopotential method. In the former case, one should demand that $h_-$ minimises the potential, while in the latter one must impose eq.~\eqref{eq:effpressure}, namely that the minima of the pseudopotential are degenerate, as appropriate for static configurations with no net pressure.

In the case of detonations the procedure outlined above simplifies: since the fluid is unperturbed in front of the wall, $T_+$ can be directly identified with $T_{\textrm{nuc}}$. It then suffices to scan only over $v_+$, which is directly related to the wall velocity. Concerning hybrids, the method is slightly more convoluted because one cannot relate the wall velocity $v_w$ to neither $v_+$ or $v_-$, so that $v_w$ is an additional parameter. However, hybrids have the additional constraint $|v_-|=c_s^-$ (see appendix \ref{sec:hydrodynamic_matching} for more details). Then, the previous procedure can be adapted as follows. As the wall velocity $v_w$ is now an additional parameter, one starts with a choice of the three parameters $v_w,v_+,T_+$. As before, knowledge of $v_+,T_+$ allows to determine all relevant quantities on both sides of the wall. Imposing the constraint $|v_-|=c_s^-$ fixes $v_+$, effectively leaving two free parameters, $T_+$ and $v_w$. One can again match the quantities at the wall to the hydrodynamic profiles for the deflagration and rarefaction waves in front of and behind the wall. The two free parameters are then fixed by imposing the same two final constraints as in the procedure for deflagrations and detonations.

\begin{figure}[t!]
    \centering
    \subfloat{
        \includegraphics[]{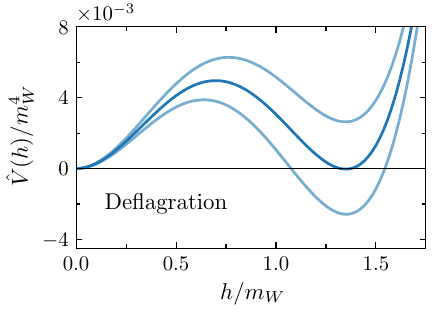}
        \captionsetup{labelformat=empty}}
    \subfloat{
        \vspace{-0.012 cm}
        \includegraphics[]{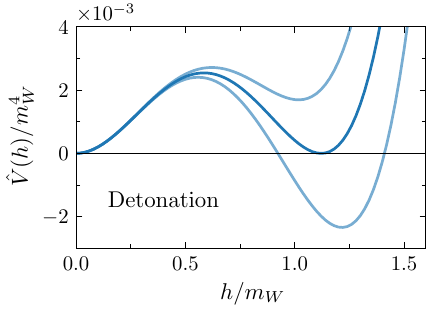}
    }
    \caption{Pseudopotential in the deflagration (left) and detonation (right) regime for $N=4$ and $\lambda_{HS} = 0.85$. The dark blue line indicates the steady state solution with $\Delta \hat{V} = 0$. For deflagrations, from top to bottom $T_+$ is $117.93$ GeV, $117.68$ GeV, $117.45$ GeV and $v_+$ is $-0.470$, $-0.444$, $-0.400$. For detonations $T_+ = T_{\text{nuc}} = 116.993$~GeV and $v_+$ is $-0.680$, $-0.735$, $-0.930$.}%
  	 \label{fig:Pseudopotential_Profile}
\end{figure}

Given the iterative procedure outlined above, the pseudopotential method offers a substantial computational advantage, particularly in the deflagration and hybrid regimes (where the matching requires a two-parameter scan over $v_+$ and $T_+$, rather than a single scan over $v_+$ as in detonations). All the relevant information about the broken-phase configuration is extracted directly from the pseudopotential, which amounts to evaluating a single integral and is therefore far more efficient than solving the scalar field equation at every step of the iteration.

For illustration, figure \ref{fig:Pseudopotential_Profile} shows the pseudopotential for $N=4,~\lambda_{HS}=0.85$ and specific configurations of  $v_+$ and $T_+$. In this regime we find a nucleation temperature $T_\text{nuc} = 116.993$~GeV and a critical temperature $T_c = 117.409$~GeV. We further obtain a unique consistent solution for both deflagrations and detonations that satisfies $\Delta \hat{V} = 0$. Notably, the behaviour of the pseudopotential differs between the deflagration and detonation regimes.
In deflagration modes, increasing $|v_+|$ leads to larger $\Delta \hat V$, i.e. an increased outward pressure, while the opposite is true for detonations. As will be seen in the next section, once the dependence on $v_+$ is traded for the wall velocity, the difference of behaviour will allow to extract conclusions about the stability of the solutions. While it might seem counter-intuitive that $\hat V(h)$ behaves differently for deflagrations and detonations, the origin lies in the structure of the temperature profile $T(h)$ obtained from the hydrodynamic equations \eqref{eq:wall_profile_equation}. These solutions are multivalued, and the deflagration and detonation constraints pick out different branches of $T(h)$, leading to the different behaviour for $\hat V(h)$.
\begin{figure}[t!]
    \centering
    \subfloat{
        \includegraphics[]{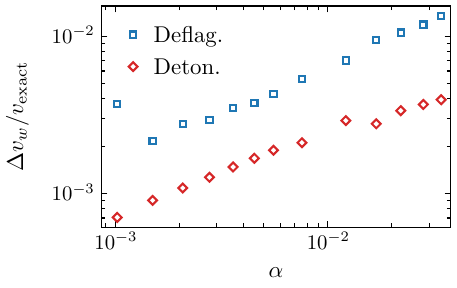}
        \captionsetup{labelformat=empty}}
    \subfloat{
        \vspace{-0.012 cm}
        \includegraphics[]{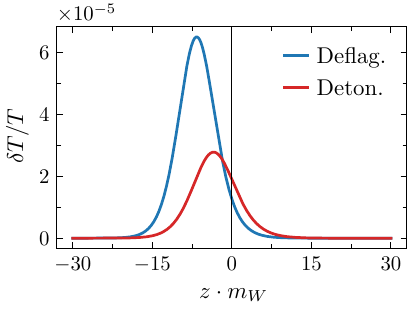}
    }
   	\caption{ Validity of taking $h^\prime(z) = 0$ in $T(h,h^\prime)$, demonstrated by explicit numerical calculation for $N=4$. The left figure shows the relative deviation between the physical wall velocity obtained from the pseudopotential and by solving eq.~\eqref{eq:differential_equation} demanding $h^{\prime \prime}(z_\text{min})= 0$. The small deviation confirms that the temperature dependence on $h^\prime$ has a negligible effect. This is illustrated in the right plot, showing the dimensionless quantity $\delta T/T$ along the full profile distance, specifically for $\lambda_{HS} = 0.85$. The results indicate that the solution obtained from the pseudopotential provides an excellent approximation to the exact solution.}
   	\label{fig:physical_vs_approx}
\end{figure}

To obtain the exact physical solutions, as an alternative to the pseudopotential method, we solve the scalar equation of motion with \texttt{Mathematica}, imposing the required boundary conditions at large $|z|$, chosen to be much larger than the typical wall width. The latter is of the order of the inverse of the typical mass-scale in the broken phase; in practice, we find that $z_\text{max} \in [12/m_W,50/m_W]$ allows to find solutions that satisfy the desired boundary conditions with good numerical accuracy. 

The left plot in figure \ref{fig:physical_vs_approx} shows the relative difference $\Delta v_w / v_\text{exact}$, with $\Delta v_w = |v_{\text{pseudo}} - v_{\text{exact}}|$, where $v_{\text{pseudo}}$ is obtained from the pseudopotential method and $v_{\text{exact}}$ from solving the scalar equation of motion, as a function of the strength of the phase transition. We quantify the latter by the phase transition strength parameter~\cite{Athron:2023xlk,Giese:2020znk}
\begin{equation}
\alpha \equiv \frac{4(\theta_+ - \theta_-)}{3 \omega},
\end{equation}
where $\theta_\pm = \tfrac{1}{4}(\rho - 3p)$ denotes the trace anomaly evaluated in the symmetric and broken phases, respectively. For simplicity, $\alpha$ is evaluated at $T = T_\text{nuc}$, such that for each parameter point both deflagration and detonation solutions are shown at the same value of $\alpha$.
The plot spans more than an order of magnitude in $\alpha$, ranging from $10^{-3}$ to $\sim 0.035$, while the accuracy of the wall velocity estimate remains at the percent level or better. Although the discrepancy increases mildly towards larger values of $\alpha$, the agreement remains very good, demonstrating that the pseudopotential approach provides accurate wall velocity predictions across the entire range.

The deviation $\Delta v_w$ is expected to originate from the $h^\prime$-dependence of the temperature, which is neglected in the pseudopotential method. To quantify this effect, we expand $T(h,h^\prime)$ around $h^\prime=0$,
\begin{align}
T(h,h^\prime) = T_0 + h^{\prime 2} \bigg( \frac{1}{2}  \left. \frac{\partial^2 T(h,h^\prime)}{\partial h^{\prime 2}} \right|_{h^\prime = 0} \bigg)  + \mathcal{O}(h^{\prime 4}),
\end{align}%
and find that the leading dependence of the temperature on the field gradient is captured by the dimensionless quantity
\begin{align}
\frac{\delta T}{T} \equiv \frac{h^{\prime 2}}{2T_0} \left. \bigg( \frac{\partial^2 T(h,h^\prime)}{\partial h^{\prime 2}} \right|_{h^\prime = 0} \bigg) .
\label{eq:deltaTTfull}
\end{align}
\begin{figure}[!t]
    \begin{center}
        \includegraphics{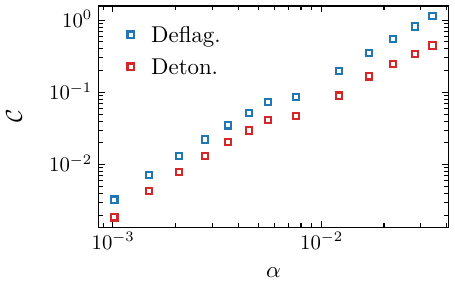}
        \caption{Parameter $\mathcal{C}$ of Eq.~\eqref{eq:DeltaphiT2Lw} as a function of the phase transition strength $\alpha$, obtained from the parameter scan of our toy model. Although increasing mildly with $\alpha$, it remains well below $\sim \mathcal{O}(10)$, validating eq.~\eqref{eq:DeltaphiT2Lw}.}
        \label{fig:DeltaphiT2Lw}
    \end{center}
\end{figure}
 
The behaviour of eq.~\eqref{eq:deltaTTfull} is illustrated in the right plot of figure \ref{fig:physical_vs_approx}, where the quantity is evaluated along the full bubble wall profile for a representative parameter point, $N=4$ and $\lambda_{HS}=0.85$. The correction is localised around the wall, where the field gradient is largest, and rapidly vanishes in the asymptotic regions. Its magnitude remains below the per mille level, consistent with the small deviations observed in the wall velocity.

To obtain an estimate of $\delta T/T$, it is convenient to derive an approximate relation for eq.~\eqref{eq:deltaTTfull} in terms of physical parameters. This can be achieved by solving the hydrodynamic equations for $T(\phi,\phi^\prime)$ perturbatively in the field gradient, as detailed in appendix~\ref{sec_appendixB}. Retaining the leading contributions, one finds
\begin{align}
\frac{\delta T}{T} \sim \frac{\Delta\phi^2}{8 a T^4 L_w^2},
\end{align}
where $a=\pi^2 g_*/90$, $\Delta\phi$ denotes the field difference and $L_w$ the wall width. For typical values $g_* \sim \mathcal{O}(10^2)$, this indicates that neglecting higher-order gradient corrections in the temperature, as required for the pseudopotential construction, is justified when
\begin{align}
 \mathcal{C} \equiv \frac{\Delta\phi}{T^2 L_w}\ll 10.
 \label{eq:DeltaphiT2Lw}
\end{align}
The former quantity turns out to be strongly correlated with the phase transition strength $\alpha$, as  shown in figure~\eqref{fig:DeltaphiT2Lw}.

\section{Results for the bubble wall velocity}
\label{sec:results}
\begin{figure}[!t]
    \begin{center}
        \includegraphics{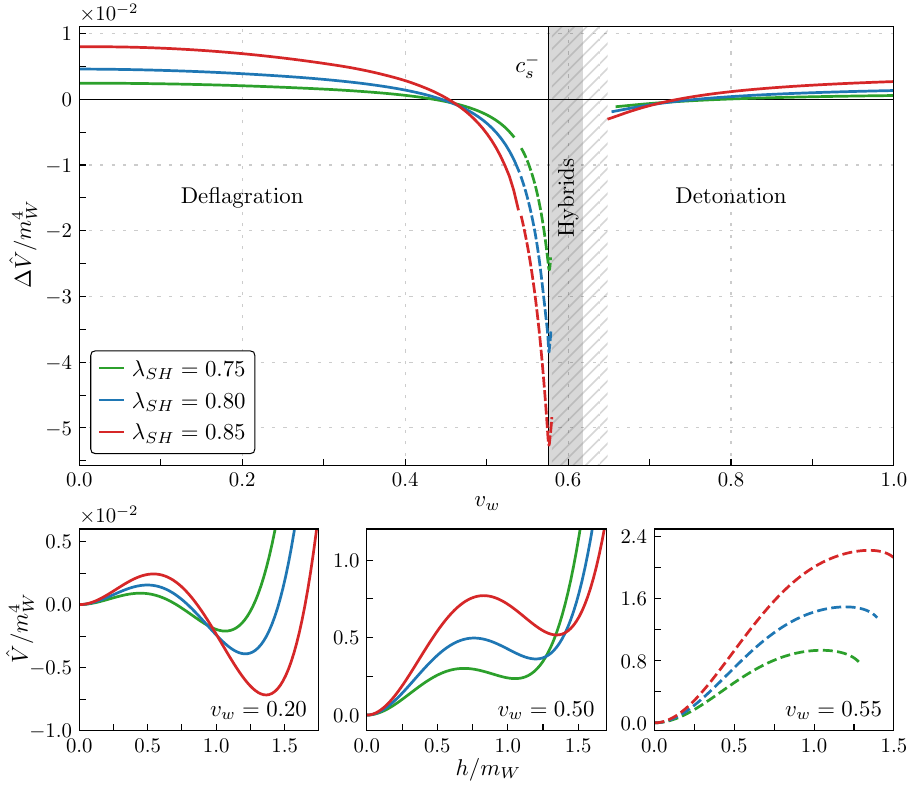}
        \caption{\textbf{Upper plot:} Effective pressure as a function of the wall velocity for three different values of the coupling constant $\lambda_{HS}$. Solid lines indicate solutions where the extrema of $\hat{V}(h)$ are separated by a barrier, while dashed lines correspond to cases without a barrier, i.e. when one of the extrema is a maximum of $\hat{V}(h)$. The grey contour area depicts the segment in which hybrid solutions are expected to exist, the light grey hatched region indicates a region in which no solutions to the hydrodynamic matching constraints could be found. The physical solutions with $\Delta \hat V=0$ for deflagrations are stable, whereas the detonation solutions are expected to decay either to the left or the right. Stationary hybrid solutions cannot be found. \textbf{Lower plot:}  Pseudopotential $\hat V(h)$ for three values of the bubble wall velocity and three choices of the coupling constant $\lambda_{HS}$.}
        \label{fig:eff_pressure_full_profile_time}
    \end{center}
\end{figure}

We now present the results for the net pressure $\Delta \hat{V}$ across the full range of wall velocities for fixed $N=4$ and three different values of the coupling constant: $\lambda_{HS} = 0.75$, $\lambda_{HS} = 0.80$ and $\lambda_{HS} = 0.85$. The corresponding nucleation temperatures are, respectively, $T_\text{nuc} = 118.707~\text{GeV}$, $T_\text{nuc} = 117.795 ~\text{GeV}$, $T_\text{nuc} = 116.993~\text{GeV}$, and the critical temperatures are $T_c = 118.915~\text{GeV}$, $T_c = 118.096~\text{GeV}$, $T_c = 117.409~\text{GeV}$. The results are shown in figure \ref{fig:eff_pressure_full_profile_time}. Static solutions were only found for detonations and deflagrations. There is no configuration of $v_+$ and $T_+$ that satisfies all the hybrid conditions along the hydrodynamic profile with $\Delta \hat{V} = 0 $. The hybrid branch exists only over a small range of wall velocities, close to $v_w = c_s$. There is also a segment between the hybrids and detonations where no solutions could be found (indicated by the light grey hatched region), as the hydrodynamic matching conditions cannot be satisfied in this regime.

The interpretation of the plot provides valuable insight into the bubble wall dynamics. $\Delta \hat{V}>0$ indicates a positive net pressure acting on the bubble wall, meaning that the driving force exceeds the backreaction force and the bubble gets accelerated. This causes the bubble wall velocity to increase, effectively pushing the solution toward higher velocities. In contrast, a negative net pressure $\Delta \hat{V}<0$ indicates that the backreaction force overcompensates the driving force, decelerating the wall and moving the solution toward lower velocities. This has consequences for the stability of the solutions that can be seen in the plot. For deflagrations, perturbations to the right of the physical solution (higher wall velocities) lead to $\Delta \hat{V}<0$, which slows down the wall and will restore the stable solution. Similarly, perturbations to the left (lower wall velocities) result in $\Delta \hat{V}<0$, accelerating the wall and pushing it back to the right. Therefore, deflagration solutions are expected to be stable with a bubble wall that expands at a constant subluminal speed. However, detonations exhibit the opposite behaviour: If the steady velocity solution is perturbed to the left, the backreaction force takes over and causes the solution to decay further towards smaller velocities, leading to a collapse of the bubble wall. Perturbations to the right result in $\Delta \hat{V}>0$, accelerating the wall until it reaches light speed. Consequently, one would expect stable solutions to exist only for deflagrations, whereas detonation solutions are inherently unstable and are expected to decay into runaway configurations.  Our results in figure \ref{fig:eff_pressure_full_profile_time} are analogous to those found in  Ref.~\cite{Laurent:2022jrs} for the net \emph{inward} pressure using calculations relying on a $\tanh$ ansatz for the field profile.

While we did not find configurations covering the full range of bubble velocities between deflagrations and detonations, we confirm that the backreaction pressure grows sharply for deflagrations with wall velocities close to the speed of sound, where it reaches its maximum. This agrees qualitatively with Ref.~\cite{Laurent:2022jrs}, although in the latter the peak occurs at the Jouguet velocity in the hybrid regime, rather than at the speed of sound. For hybrids and detonations we find the backreaction pressure to decrease with the velocity.  

Hybrid configurations require enforcing the condition $|v_-| = c_s^-$ behind the wall. In our implementation $v_-$ is determined from the position $h_-$ of the broken-phase extremum of the pseudopotential $\hat V(h)$. When approaching the hybrid regime, for large regions of parameter space we find that this extremum disappears before the condition $|v_-| = c_s^-$ can be reached. This occurs because, for the corresponding values of $v_+$ and $T_+$, the hydrodynamic equations cease to admit real solutions for sufficiently large field values. As a consequence of this the pseudopotential can only be defined up to a maximal field value and no longer develops a second extremum corresponding to the broken phase. This means that the hydrodynamic equations are incompatible with configurations interpolating between energy minima and satisfying $|v_-| = c_s^-$ --- in other words, for such parameter choices hybrid bubbles are not allowed. This absence of solutions between hybrid and detonation configurations has been observed in other studies \cite{Krajewski:2023clt,Krajewski:2024zxg} and  explains the gap near $v_w=c_s^-$ in Fig.~\ref{fig:eff_pressure_full_profile_time}. For $v_w$ very close to $c_s^-$ we do find hybrid solutions over a very narrow range of wall velocities. In this case it turns out that the pseudopotential is defined for field values up to the extremum $\phi_-$ leading to $v_-=c_s^-$, and not beyond.
In practice this means that the hybrid condition can only be approached as a limiting case in our scan over $v_+$ and $T_+$. To probe this regime we allow for a small tolerance and consider configurations satisfying $c_s^- - |v_-| < 10^{-5}$.

The behaviour of the backreaction force in Fig.~\ref{fig:eff_pressure_full_profile_time} suggests that a slowly accelerating bubble wall will only lead to deflagration solutions as it arrives at the steady state solution before it can reach the detonation branch. Additionally, the sharp growth of the backreaction force close to $v_w=c_s$ results in terminal velocities for deflagrations not far from $c_s$, and with little variation with the interaction strength between the scalar field in the plasma --- in our results in figure \ref{fig:eff_pressure_full_profile_time}, we consistently predict $v_w\sim 0.45$. The conclusion about the absence of detonation solutions with a constant velocity matches the discussion and results in Refs.~\cite{Ai:2023see,Krajewski:2024gma}. We note, however, that the latter reference, which used numerical simulations to study the time-evolution of bubble configurations in the plasma, found, that the steady-state deflagration solutions were not always reached. It remains to understand how this is compatible with the  growth of the backreaction force for $v_w=c_s$.

\section{Conclusions}
\label{sec:conclusion}
In this work, we have introduced a novel method to estimate terminal bubble velocities during first-order phase transitions in a plasma that remains in LTE. Its main advantage is that it provides a direct physical description of the dynamics in terms of the net outward pressure acting on the bubble wall, which also makes it useful beyond the mere determination of stationary solutions. From a practical point of view, the method furthermore bypasses the need to solve the scalar equation of motion explicitly, while requiring neither a simplified parametrisation of the fluid stress-energy tensor nor a particular ansatz for the field profile.

The method relies on a function of the scalar field that we refer to as the ``pseudopotential'' $\hat V(\phi)$. Its most relevant properties are that its shape changes as a function of the wall velocity, while its extrema match those of the ordinary potential at a fixed temperature. Moreover, the  difference between the values of the pseudopotential at two extrema can be identified with the net outward pressure acting on a bubble wall in which the scalar field interpolates between two minima of the ordinary potential. Hence, the terminal wall velocity of bubbles can be determined from requiring a null net pressure, which corresponds to degenerate minima in the pseudopotential.

Aside from the assumption of LTE --- which is expected to lead to upper bounds on the wall velocity --- the method relies on the following approximation: when using the hydrodynamic equations to solve for the temperature as a function of the scalar background and its derivatives, one can neglect the dependence on the latter: $T(\phi,\phi')\approx T(\phi)$. We have quantified the validity of this approximation by deriving the constraint $\mathcal{C} = \Delta \phi / (T^2 L_w) \ll  \mathcal{O}(10)$, which ensures that gradient corrections to the temperature profile can be neglected. We have tested this condition in an extension of the SM with complex scalars featuring a first-order electroweak phase transition, and we confirmed that it is satisfied across the parameter space probed in our work, which features phase transition strengths in the range $\alpha \sim 10^{-3}$ up to $\alpha \sim 0.035$. Within this regime, we find that the pseudopotential method allows us to estimate wall velocities in LTE with an accuracy of $\sim 1\%$ or better compared to the values obtained by solving the scalar equation of motion.

The pseudopotential method, supplemented with the appropriate matching with deflagration, detonation or hybrid hydrodynamic profiles away from the bubble, allows for an efficient determination of wall velocities for bubbles satisfying all the required boundary conditions. Looking for configurations that interpolate between minima of $\hat{V(\phi)}$ across the wall, without imposing the condition $\Delta\hat{V}=0$ for stationary bubbles, we found one-parameter branches of solutions for deflagrations, detonations and hybrids. Choosing the wall velocity as the relevant parameter, our main results are shown in figure \ref{fig:eff_pressure_full_profile_time}. The slope of the net outward pressure near the stationary $\Delta\hat{V}=0$ solutions shows that, while stationary deflagrations are stable, the corresponding detonations are unstable, in keeping with similar conclusions in the literature \cite{Krajewski:2024gma,Ai:2023see,Laurent:2022jrs}. Interestingly, we found no stationary hybrid configurations.  Although this may reflect the restricted parameter space explored here, it would be worthwhile to investigate whether such solutions arise in broader regions of theory space.

Finally, it would be interesting to study whether the pseudopotential method can be extended beyond the LTE regime. This would in principle be possible for corrections to the equation of motion that can be expressed as functions of the scalar field, and not of its derivatives. Note that this applies to part of the terms involving condensate-induced interactions found in Ref.~\cite{Ai:2025bjw}, as in a local approximation they can be expressed as a sum of mass corrections  plus gradient-dependent contributions; the former could be incorporated into the pseudopotential.

\section*{Acknowledgments}
The authors acknowledge support from the Cluster of Excellence ``Precision Physics, Fundamental Interactions, and
Structure of Matter'' (PRISMA+ EXC 2118/1) funded by the Deutsche Forschungsgemeinschaft (DFG, German Research Foundation) within the German Excellence Strategy
(Project No. 390831469).

\begin{appendix}

\section{Hydrodynamic matching}
\label{sec:hydrodynamic_matching}
\subsection{Detonations}

\begin{figure}[!t]
\centering
\begin{normalsize}
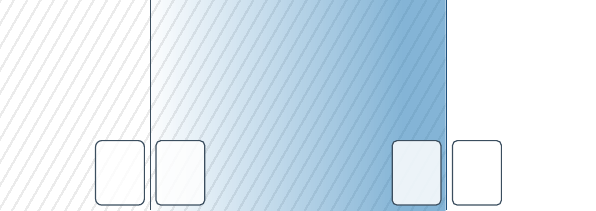
\end{normalsize}
\caption{Illustration of the geometry of a detonation. Red arrows indicate velocities measured in the wall-/rarefaction-front frames, black arrows represent velocities observed in the rest frame of the bubble centre. Dashed contour lines indicate the broken phase. The black vertical lines indicate the bubble wall/rarefaction front. The rarefaction front only exists in a planar approximation.}
\label{fig:detonation_qualitative}
\end{figure}

A qualitative illustration of the detonation profile is shown in figure \ref{fig:detonation_qualitative}. This graphic primarily serves as a visual tool to help clarify the relationship between the different velocities, making it easier to keep track of the dynamics in each reference frame. From now on we use subscripts $+$ and $-$ to denote quantities that are measured in the reference frame of the bubble wall: The $+$ represents quantities in the symmetric state in front of the wall, while the $-$ denotes quantities behind the wall. A similar notation is introduced for the quantities along the rarefaction front, but with a hat on top.

In detonations, the phase transition wall moves with $\xi_w > c_s$ and collides with the undisturbed fluid in front of the wall. As the fluid in front is at rest, one can directly identify the wall velocity with $\xi_w = -v_+$. We can also directly identify the temperature in front of the wall with the nucleation temperature $T_+ = T_\text{nuc}$, which is a known quantity. At the bubble wall front, using $\phi_+=0$ corresponding to the symmetric phase, this leaves 4 unknowns $\phi_-, T_-,v_-,\xi_w$. The scalar equation of motion \eqref{eq:differential_equation}, plus the additional requirement of $\phi_-$ being a minimum of the potential, and the plasma equations \eqref{eq:hydro_diff_equation1}~and~\eqref{eq:hydro_diff_equation2}  effectively give 4 constraints that allow to fix all the parameters related with the wall. When using the pseudopotential method, the role of the constraints related to the scalar field are played by the requirements of $\phi_-$ being a minimum of the pseudopotential, and the latter having degenerate minima. In practice, one can proceed by choosing a value of $\xi_w$, and solve \eqref{eq:differential_equation} by imposing  $\phi(z) \rightarrow 0, z \to \infty $,  $\phi^{\prime}(z)\rightarrow 0, z \to \pm \infty$, for which solutions can be found if $V(h,T(h,h^\prime))$ in the broken phase minimum is lower than in  the symmetric phase. Then one can scan iteratively over values of $\xi_w$ until the remaining boundary condition  $\phi^{\prime \prime}(z) \rightarrow 0, ~ z \to -\infty$ is satisfied. With the pseudopotential method, one can start with a value of $\xi_w$, calculate the energy difference between the minima of the pseudopotential, and scan over $\xi_w$ until the minima satisfy the desired degeneracy condition. In this approach $\phi_-$ can be identified with the location of one of the minima. Once $\phi_+,\phi_-$ are known with either method, the hydrodynamic quantities $v_-,T_-$ behind the wall follow from eqs.~\eqref{eq:wall_profile_equation} applied behind the wall, where $\phi=\phi_-,\phi'=0$.

Behind the wall the fluid enters the broken phase, where in the wall frame it slows down to $|v_-| <| v_+|$. Far behind the wall, the temperature and velocity profiles can be matched to self-similar solutions of the hydrodynamic equations \eqref{eq:hydro_diff_equation1}~and~\eqref{eq:hydro_diff_equation2}. This requires to translate quantities from the wall frame to the rest frame of the bubble centre, or fluid frame, in which the plasma is at rest far away from the wall. The fluid velocity right behind the bubble wall in this frame, denoted as $v_{\text{fluid},-}$, can be determined by a Lorentz boost involving the wall velocity $v_w=-v_+$:
\begin{align}
v_{\text{fluid},-} = \frac{v_- - v_+}{1-v_- v_+}.
\end{align}
Moving to $\xi<\xi_w$, in the solutions to \eqref{eq:hydro_diff_equation1}~and~\eqref{eq:hydro_diff_equation2} the fluid velocity slows down and drops to zero smoothly at $\xi = c_s$ \cite{Espinosa:2010hh}. Regarding the temperature, for $\xi<\xi_w$ the plasma gets colder until it reaches a constant temperature value $\hat{T}_-$ behind the rarefaction wave. From imposing consistency along the rarefaction wave it also follows that there cannot exist strong detonations ($v_- < c_s$), only weak detonations ($v_- > c_s$) and Jouguet detonations ($v_- = c_s$) \cite{Espinosa:2010hh}. If one insists on a planar approximation for the fluid profiles away from the bubble --- which corresponds to ignoring the $v/\xi$ term in eq.~\eqref{eq:hydro_diff_equation1} --- the velocity and temperature are constant except for a discontinuity at a rarefaction front in which the velocity drops to zero. While in this approximation $T$ and $v$ are discontinuous at the front, one still has continuity relations for the stress-energy tensor that follow from considering eqs.~\eqref{eq:wall_profile_equation} on both sides of the front.

\subsection{Deflagrations}
\label{sec:deflagration}
\begin{figure}[!t]
\centering
\begin{normalsize}
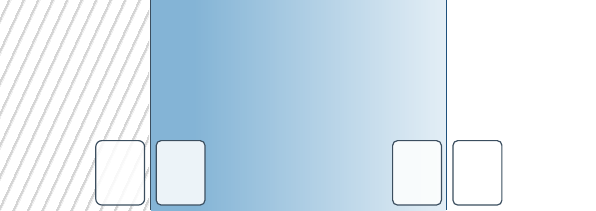
\end{normalsize}
\caption{Illustration of the geometry of a deflagration with a shock front preceding the bubble wall. Red arrows indicate velocities measured in the wall-/shock-front frames, black arrows represent velocities observed in the rest frame of the bubble centre. At the shock front the velocity jumps to zero creating a discontinuity; also beyond the planar approximation. Behind the wall the fluid stays unperturbed such that the wall velocity can be determined by $v_w = -v_-$. }
\label{fig:deflagration_qualitative}
\end{figure}

The deflagration profile is depicted in figure \ref{fig:deflagration_qualitative}. Deflagration bubble walls move with subsonic speed $\xi_w < c_s$. Unlike in detonations, the fluid is undisturbed behind the wall and the wall velocity is directly identified with $\xi_w = - v_-$. In front of the wall the fluid velocity jumps to some finite value $v_{\text{fluid},+}$ such that $|v_-| > |v_+ |$. The fluid velocity is connected to $v_+$ and $v_-$ by a Lorentz transformation
\begin{align}
v_{\text{fluid},+} = \frac{v_+ - v_-}{1-v_+ v_-}.
\label{eq:lorentz_deflag}
\end{align}
Along the shock wave of the perturbed fluid ahead of the wall, $v_{\text{fluid}}(\xi)$ decreases,  either dropping to zero smoothly or at a discontinuous shock front which can be present even when going beyond the planar approximation in eqs.~\eqref{eq:hydro_diff_equation1} and \eqref{eq:hydro_diff_equation2}, in contrast to the case of detonations. Again, conservation of the stress-energy-momentum tensor lets us match both sides of the shock front. Quantities measured in the reference frame of the shock front are denoted by a tilde. Similarly to the case of the wall frame, we adopt the convention of using subscripts $+$ and $-$ for quantities in front of and behind the shock front. The matching relations follow from eqs.~\eqref{eq:wall_profile_equation} and read
\begin{align}
\label{eq:deflag_shock_condition1}
\omega (\tilde{T}_+) \gamma^2(\tilde{v}_+) \tilde{v}_+^2 + p(\tilde{T}_+) &= \omega(\tilde{T}_-) \gamma^2(\tilde{v}_-) \tilde{v}_-^2 + p(\tilde{T}_-) \ , \\
\label{eq:deflag_shock_condition2}
\omega(\tilde{T}_+) \gamma^2(\tilde{v}_+) \tilde{v}_+ &= \omega(\tilde{T}_-) \gamma^2(\tilde{v}_-) \tilde{v}_- \ .
\end{align}
Here we made use of the fact that in the unbroken phase the scalar field has a constant background value  $\phi = 0$. 

Solving the equations allows us to eliminate two of the unknowns among $v_\pm,T_\pm,\tilde v_\pm, \tilde T_\pm$. For example, a common approximation in the literature is the bag model \cite{Espinosa:2010hh}, where the potential is approximately  $V \sim a T^4$ and $\omega = -4 a T^4$. In this case eqs.~\eqref{eq:deflag_shock_condition1}~and~\eqref{eq:deflag_shock_condition2} will lead to inward and outward fluid velocities satisfying $\tilde{v}_+ \tilde{v}_-=1/3$ in the shock-front rest frame. This simplified condition always applies when there is no field-independent mass term appearing in the potential in the broken phase; in our treatment, however, we have such terms, which involve the Higgs and singlet mass parameters $m^2_H$ and $m^2_\Phi$ (see eqs.~\eqref{eq:wall_profile_equation}), and thus we go beyond this simplification. Regardless of whether such an approximation is used or not, boosting the relation between $\tilde v_+$ and $\tilde v_-$ to the rest frame of the bubble centre gives a condition between the fluid velocity $v_{\textrm{fluid}}$ and the location $\xi_s$ of the shock front, which can be used to determine the latter.

In deflagrations, the nucleation temperature is now identified with the temperature in front of the shock front $T_\text{nuc} = \tilde{T}_+ $. The moving fluid in front of the wall leads to compression, which increases the temperature in the region ahead of the wall. Consequently, in deflagrations the temperatures always satisfy $T_+ > \tilde{T}_- > T_\text{nuc}$. Unlike in detonations, $T_+$ does not coincide with $T_\text{nuc}$. Regarding the values of the scalar field, since in front of the bubble wall the field is in the symmetric phase $\phi=0$ everywhere, there is only one unknown $\phi_-$.  Hence one can start with $10$ unknown parameters: $T_\pm, v_+,\xi_w,\phi_-,
\tilde T_-, \xi_s,\tilde v_-$ and the two values of the fluid velocity at the wall and the shock front (see figure \ref{fig:deflagration_qualitative}). These 10 parameters can be determined via   $10$ constraints: the two static plasma equations at the wall \eqref{eq:wall_profile_equation}, the static scalar equation of motion at the wall \eqref{eq:scalar_eom_planar}, the condition that $\phi_-$ is at a minimum of the potential, the two stress-energy matching conditions at the shock front \eqref{eq:deflag_shock_condition1} and \eqref{eq:deflag_shock_condition2}, two boost relations linking the fluid frame velocity $v_{\textrm{fluid}}$ with the incoming/outgoing velocities in the wall and shock frames, and the two hydrodynamic equations \eqref{eq:hydro_diff_equation1} and \eqref{eq:hydro_diff_equation2} for the fluid in between the wall and the shock front. When using the pseudopotential method, again the two constraints associated with the scalar field are replaced by the requirement of $\phi_-$ being a minimum of the pseudopotential, and the latter having degenerate minima. 

In practice, one can proceed as follows. First, one chooses two values of $T_+,v_+$. Then, one  solves the scalar equation of motion \eqref{eq:scalar_eom_planar} with the boundary conditions $h(z) \rightarrow 0, z \to \infty $,  $h^{\prime}(z)\rightarrow 0, z \to \pm \infty$ predicting $\phi_-$. The scalar equation only admits a solution if the free energy difference is positive, namely when the broken phase of $V(h,T(h,h^\prime))$ is lower in free energy than the symmetric one. Alternatively, one can compute the pseudopotential and identify $\phi_-$ with the location of the second minimum. Once $\phi_-$ is known, $T_-,\xi_w=-v_-$ follow from eqs.~\eqref{eq:wall_profile_equation}. With $v_+,v_-$ known, the fluid velocity in front of the wall follows from the boost relation of eq.~\eqref{eq:lorentz_deflag}. Then one can compute the hydrodynamic profile in front of the wall by solving the hydrodynamic equations \eqref{eq:hydro_diff_equation1}, \eqref{eq:hydro_diff_equation2} with the appropriate boundary conditions for $v_{\textrm{fluid}}$ and $T=T_+$ ahead of the wall. If the velocity profile does not smoothly reach zero, a shock front must be present, whose location is found by demanding compatibility with the matching equations \eqref{eq:deflag_shock_condition1}, \eqref{eq:deflag_shock_condition2} as explained above. From these matching equations one predicts the value $\hat T_+$ of the temperature ahead of the front. 

To summarise, one starts choosing $v_+,T_+$ and predicting all the other quantities, in particular $\phi_-$ and $\hat{T}_+$. Then one can scan over $T_+,v_+$ until the two remaining constraints are satisfied, namely $\hat{T}_+=T_{\textrm{nuc}}$ and, depending on whether one is solving for the equation of motion of the scalar or using the pseudopotential method, one of the two conditions that follow. In the first case, one should require that $\phi_-$ is at a minimum of the potential, while in the second case, one should impose degeneracy between the two minima of the pseudopotential.

\subsection{Hybrids}
\begin{figure}[t!]
\centering
\begin{normalsize}
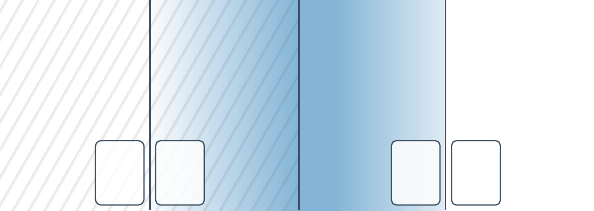
\end{normalsize}
\caption{Illustration of the geometry of a hybrid profile that includes both detonation and deflagration solutions. Red arrows indicate velocities measured in the wall-/shock- or rarefaction-front frames, black arrows represent velocities observed in the rest frame of the bubble centre. The rarefaction wave follows the bubble wall with a relative velocity $|v_-|= c_s^-$. }
\label{fig:hybrids_qualitative}
\end{figure}

For completeness we consider hybrids, which combine both deflagration and detonation solutions into one combined solution. The graphic in figure \ref{fig:hybrids_qualitative} shows the fluid velocity profile with all the unknowns. The bubble wall is preceded by a shock front that moves with velocity $\xi_s$ and is followed by a rarefaction front with velocity $\xi_r$. Applying the hydrodynamic constraints for deflagration and detonation solutions yield $|v_-| \ge c_s^-$ and $|v_+| \le c_s^+$. However, strong deflagrations with $|v_-| > c_s^-$ are known to be unstable and tend to split into other configurations \cite{LANDAU1987505}. This has also been confirmed by numerical calculations in \cite{Kurki-Suonio:1995rrv}. Therefore we get an additional constraint by $|v_-| \le c_s^-$, which ensures that the fluid behind the bubble wall moves relative to the wall with the speed of sound:
\begin{align}\label{eq:vminus_cs}
|v_-| = c_s^-
\end{align}
This requirement is essential to find consistent solutions, since now $\xi_w$ cannot be directly identified with neither $v_-$ nor $v_+$ due to the fluid being perturbed on both sides of the wall. Consequently,  $\xi_w$ introduces a new independent parameter requiring an additional constraint (provided by eq.~\eqref{eq:vminus_cs}) to fully resolve the system. The remaining quantities are then determined following a procedure analogous to that used for deflagrations and detonations.

\section{Field gradient effects on the temperature}
\label{sec_appendixB}

The temperature profile along the wall follows from the hydrodynamic equations~\eqref{eq:wall_profile_equation}, which determine $T$ as a function of $\phi$ and its gradient $\phi'$. Solving the second equation for the velocity $v(\phi,T)$ and inserting the result into the first yields a single implicit equation for the temperature,
\begin{align}
F(T,\phi,\phi^\prime) \equiv c_2 \, v(\phi,T) + \tfrac{1}{2}( \phi^\prime)^2 - V(\phi,T) - c_1 = 0 .
\label{eq:Ffunction_app}
\end{align}

To quantify the dependence of the temperature profile $T(\phi,\phi^\prime)$ on the field gradient, we expand around $\phi^\prime=0$. Since $F(T,\phi,\phi^\prime)$ is even in $\phi^\prime$, the temperature must satisfy $T(\phi,\phi^\prime)=T(\phi,-\phi^\prime)$, implying the absence of a linear term. Introducing $\chi=(\phi^\prime)^2$, we write
\begin{align}
T(\phi,\chi)= T(\phi,0) + \chi \; \frac{\partial T}{\partial \chi} \bigg{|}_{\chi=0} + \mathcal{O}(\chi^2).
\label{eq:T_expansion}
\end{align}
The coefficient $T_2 \equiv \partial_\chi T|_{\chi=0}$ is obtained by differentiating eq.~\eqref{eq:Ffunction_app} with respect to $\chi$,
\begin{align}
\frac{\mathrm{d} F}{\mathrm{d} \chi} \bigg{|}_{\chi=0} 
= \frac{\partial F}{\partial T} \frac{\partial T}{\partial \chi}\bigg{|}_{\chi=0} 
+ \frac{\partial  F}{\partial \chi}\bigg{|}_{\chi=0} = 0,
\end{align}
which gives
\begin{align}
T_2 = \frac{\partial T}{\partial \chi}\bigg{|}_{\chi=0} 
= - \frac{\partial_\chi F}{\partial_T F} \bigg{|}_{\chi=0}.
\end{align}
Evaluating the derivatives from eq.~\eqref{eq:Ffunction_app} yields
\begin{align}
T_2 = \frac{1}{2}\frac{1}{\partial_T V(\phi,T_0) - c_2 \, \partial_T v(\phi,T_0)},
\end{align}
where $T_0 \equiv T(\phi,0)$ denotes the temperature profile in the absence of gradient corrections. The leading correction is therefore
\begin{align}
\frac{\delta T}{T_0} \sim \frac{(\phi^\prime)^2}{2T_0 \,(\partial_T V(\phi,T_0) - c_2 \, \partial_T v(\phi,T_0))}.
\label{eq:deltaT_general}
\end{align}

The second term in the denominator can be expressed in terms of thermodynamic quantities. Using the solution for the velocity,
\begin{align}
v(\phi,T) = \frac{-1 \pm \sqrt{1+4 X^2}}{2 X}, \qquad X = \frac{c_2}{\omega},
\end{align}
its temperature derivative can be written as
\begin{align}
\frac{\partial v}{\partial T}
= \frac{\partial v}{\partial X}\frac{\partial X}{\partial T}
= \frac{\partial v}{\partial X} \left( - \frac{c_2}{\omega^2} \frac{\partial \omega}{\partial T} \right).
\end{align}
To estimate the magnitude of eq.~\eqref{eq:deltaT_general}, it is sufficient to retain the dominant temperature-dependent contribution to the pressure, namely the radiation term,
\begin{align}
p(T) \simeq a T^4, 
\qquad a=\frac{\pi^2 g^\star}{90},
\label{eq:t4pressure}
\end{align}
which implies
\begin{align}
\partial_T p = 4 a T^3, 
\qquad \partial_T \omega = 4 \frac{\omega}{T}.
\end{align}
Combining these relations, one finds
\begin{align}
c_2 \, \partial_T v(\phi,T_0)
= 2 \, \partial_T V(\phi,T_0)
\left( 1 \mp \frac{1}{\sqrt{4X^2 + 1}} \right).
\end{align}
Substituting into eq.~\eqref{eq:deltaT_general} leads to
\begin{align}
\frac{\delta T}{T_0} 
\sim \frac{(\phi^\prime)^2}{2T_0 \, \partial_T V(\phi,T_0) (Y - 1)}
\approx \frac{(\phi^\prime)^2}{8 a T_0^4 (Y - 1)},
\label{eq:deltaT_final}
\end{align}
with $Y= \tfrac{2}{\sqrt{4(\gamma^2 v)^2 + 1}}$. 

\begin{figure}[t!]
    \centering
    \includegraphics[width=0.92\linewidth]{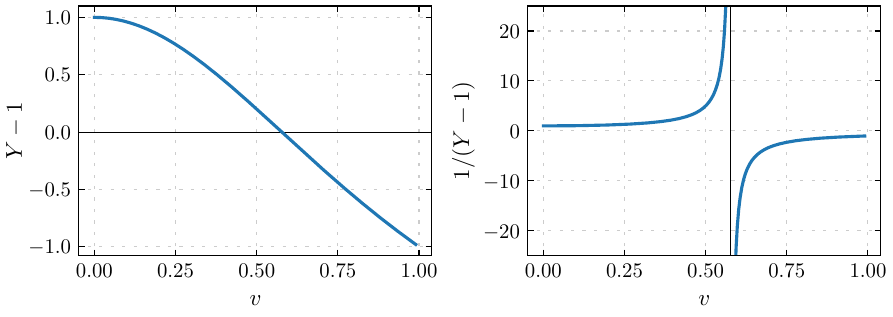}
   \caption{$Y-1$ and its inverse as functions of the velocity $v$, using only the dominant temperature contribution in eq.~\eqref{eq:t4pressure}. In the deflagration and detonation regimes, the inverse remains of order $\mathcal{O}(1)$, while it develops a singularity as $v \to c_s$.}
    \label{fig:Yminus1}
\end{figure}

The behaviour of the factor $Y-1$ is shown in fig.~\eqref{fig:Yminus1}. It remains of order unity for both deflagration and detonation solutions. Although its inverse formally develops a singularity at $Y = 1$, this is not expected to lead to a physical enhancement of the temperature correction.  The limit $Y \to 1$ corresponds to $v \to c_s$ and is associated with the hybrid regime, where the fluid velocity behind the wall approaches the speed of sound. However, this condition is only reached asymptotically in the fluid profile behind the wall~\cite{Espinosa:2010hh}. In this region the scalar field approaches its broken-phase value, $\phi \to \phi_-$, and the field gradient is therefore expected to vanish, $\phi' \to 0$. The quantity $Y-1$ approaches zero linearly in the velocity as $v\rightarrow c_s$. Eq.~\eqref{eq:Ffunction_app} suggests that $\phi'^2$ will also approach its asymptotic value of zero linearly in the velocity, leading to a finite limit of $(\phi')^2/(Y-1)$.

In view of the above, we can estimate the size of $\delta T/T_0$ away from the $Y\rightarrow1,\phi'\rightarrow0$ regime by taking $Y\sim 1$ and estimating the field gradient in terms of the characteristic length scale of the wall,  set by its width $L_w$, over which the field interpolates between $\phi_-$ and $\phi_+$. This implies
\begin{align}
\phi^\prime \sim \frac{\Delta \phi}{L_w}
\end{align}
with $\Delta\phi\equiv\phi_+-\phi-$. With the former approximations, we obtain a simple estimate for the size of gradient-induced corrections to the temperature profile:
\begin{align}
\frac{\delta T}{T_0} \sim \frac{\Delta \phi ^2}{8 a T_0^4 L_w^2} .\label{eq_18}
\end{align}

\end{appendix}

\bibliographystyle{osajnl}
\bibliography{lib}{}

\end{document}